\documentclass[twocolumn]{aastex631} 
\usepackage[utf8]{inputenc}
\usepackage{newunicodechar,graphicx}

\DeclareRobustCommand{\okina}{%
  \raisebox{\dimexpr\fontcharht\font`A-\height}{%
    \scalebox{0.8}{`}%
  }%
}
\newunicodechar{ʻ}{\okina}

\begin{document}

\title{A Deep Search for Exomoons Around WISE 0855 With JWST}

\author[0000-0003-3008-1975]{Mikayla J. Wilson}
\altaffiliation{NSF Graduate Research Fellow}
\affiliation{Department of Astronomy and Astrophysics, University of California, Santa Cruz, Santa Cruz, CA 95064, USA}

\author[0000-0002-9521-9798]{Mary Anne Limbach}
\affiliation{Department of Astronomy, University of Michigan, Ann Arbor, MI 48109, USA}

\author[0000-0001-6098-3924]{Andrew J. Skemer}
\affiliation{Department of Astronomy and Astrophysics, University of California, Santa Cruz, Santa Cruz, CA 95064, USA}

\author[0000-0003-0489-1528]{Johanna M. Vos}
\affiliation{School of Physics, Trinity College Dublin, The University of Dublin, Dublin 2, Ireland}
\affiliation{Department of Astrophysics, American Museum of Natural History, Central Park West at 79th Street, New York, NY 10024, USA}

\author[0000-0002-5500-4602]{Brittany E. Miles}
\affiliation{Steward Observatory, University of Arizona, Tucson, AZ 85721, USA}

\author[0000-0003-4225-6314]{Melanie J. Rowland}
\affiliation{University of Texas at Austin, Department of Astronomy, 2515 Speedway C1400, Austin, TX 78712, USA}

\author[0000-0001-7246-5438]{Andrew Vanderburg}
\affiliation{Department of Physics and Kavli Institute for Astrophysics and Space Research, Massachusetts Institute of Technology, Cambridge, MA 02139, USA}

\author[0000-0002-6294-5937]{Adam C. Schneider}
\affiliation{United States Naval Observatory, Flagstaff Station, 10391 West Naval Observatory Road, Flagstaff, AZ 86005}

\author[0000-0002-4404-0456]{Caroline Morley}
\affiliation{Department of Astronomy, The University of Texas at Austin, Austin, TX 78712, USA}

\author[0009-0008-5864-9415]{Brooke Kotten}
\affiliation{Department of Astronomy, University of Michigan, Ann Arbor, MI 48109, USA}

\author[0009-0004-7103-7828]{Andrew Householder}
\affiliation{Department of Astronomy, University of Michigan, Ann Arbor, MI 48109, USA}

\author[0000-0003-3444-5908]{Roxana Lupu}
\affiliation{Eureka Scientific, Inc., Oakland, CA 94602}

\author[0000-0001-5864-9599]{James Mang}
\altaffiliation{NSF Graduate Research Fellow}
\affiliation{Department of Astronomy, University of Texas at Austin, Austin, TX 78712, USA}

\author[0000-0001-9333-4306]{Richard Freedman}
\affiliation{SETI Institute, Mountain View, CA, USA}



\begin{abstract}
JWST is collecting time-series observations of many free-floating planets (FFPs) to study their weather, but these light curves are the ideal datasets to search for exomoons that transit the FFP during observations. In this paper, we present observations of the planetary-mass Y dwarf ($T=250-285K$, $M = 6.5\pm3.5 M_{Jup}$, d = 2.3\,pc) WISE J085510.83-071442.5 (WISE 0855), whose proximity and brightness make it ideal for a transiting exomoon search. We examine 11 hours of time-series spectra from the JWST Near-Infrared Spectrograph (NIRSpec) whose sensitivity, in combination with Gaussian process (GP) modeling, allows for the disentanglement of exomoon transits from WISE 0855's variability. We do not find statistically significant evidence of an exomoon transit in this dataset. 
Using injection and recovery tests of artificial transits for depths ranging between 0.1-1\% (0.35-1.12 $R_{\oplus}$) we explore the exomoon parameter space where we could successfully detect transits. For transit depths $\geq 0.5\%$ (1.96\,$R_{\text{Titan}}$), our detection rate is 96\%, which, for WISE 0855, corresponds to a moon with a companion-to-host mass ratio similar to that of Titan and Saturn. Given our sensitivity, transit probabilities, and our observational duration, we determine a $\sim$91\% probability of detecting a Titan mass analog exomoon after 18 such observations if every observed system hosts a Titan mass analog exomoon in a Galilean-like system. This suggests that \textit{JWST} observations of dozens of FFPs could yield meaningful constraints on the occurrence rate of exomoons. This paper is the first demonstration that JWST is sensitive to Galilean moon mass analogs around FFPs. 
\end{abstract}




\section{Introduction}
\subsection{Exomoons}

With our own Solar System hosting just under 300 moons, and based on the outcomes of satellite formation models, we expect satellites to be a common outcome of giant planet formation \citep{canup2006, cilibrasi2018, inderbitzi2020, cilibrasi2021}. While moons can form within a circumplanetary disk, they can also originate from satellite capture, like in the case of Neptune's retrograde moon Triton, or through giant impacts, as with the origin of Earth's Moon \citep{canup2001, agnor2006}.  Moons with diverse environments can serve as some of the most interesting sites in the search for life within our Solar System. For example, Jupiter's Europa and Saturn's Enceladus harbor liquid oceans beneath icy surfaces with various bioessential elements that have the potential to sustain life \citep{porco2006, hand2007, vance2023, davila2024}. 

The moons that formed in circumplanetary disks around our Solar System's gas-giants appear to share a common mass ratio: the total mass of moons around each planet is $\sim$0.01\% of the planet's mass. Using models of satellite formation, \citet{canup2006} find that this may be regulated by a balance between inflowing material and satellite loss through orbital decay.  However, the universality of this theory is observationally untested beyond the 4 giant planets in our own Solar System.



Moons outside of our Solar System, referred to as exomoons, could provide varied examples of moon formation in larger samples and in environments that differ from our own Solar System.  In this relatively new field, despite numerous studies and various approaches, we have yet to confirm an exomoon detection. Some examples of methodologies discussed in the literature include transit timing variations \citep{teachey2018}, transit duration variations \citep{kipping2009}, direct imaging \citep{Limbach2013}, spectroastrometry \citep{agol2015}, radial velocity searches \citep{ruffio2023}, radio emission \citep{noyola2016}, and microlensing \citep{liebig2010, fu2025}.

As described in \citet{Limbach2021}, we can also search for moon transit signatures within a planet's light curve. In addition to bound, wide-orbit exoplanets, this technique works with free-floating planets (FFPs), which circumvents the considerable issue of host starlight drowning out transit signals. The FFP population is likely made up of a combination of ejected planets that formed around stars and FFPs that formed as a very low-mass outcome of a star-forming region \citep{miretroig2023}. Satellites could probably be found around either of these FFP populations, although they might look slightly different since not all moons, especially those on wide orbits, survive ejections \citep{rabago2018, hong2018}. Since the mass range of FFPs is similar to that of solar system giant planets, discovering this population would allow us to understand how they compare to moons in our own solar system. We use this approach to search for transiting exomoon signatures within \textit{James Webb Space Telescope} (JWST) time-series observations of the free-floating planetary mass object, WISE 0855. 

\subsection{WISE 0855}
At 2.3 pc, WISE 0855 is one of our closest neighbors \citep{Luhman2014}. This isolated body is the coldest known Y dwarf, of which there are only $\sim$50, with a temperature of 250-285K \citep{Luhman2024, rowland2024, kuhnle2025}. As the coldest brown dwarf, it spectroscopically resembles giant planets, including Jupiter at some IR wavelengths, and has a WISE W2 band (4.6 $
\mu$m) magnitude of 14.016 $\pm$ 0.048
\citep{skemer2016, Luhman2016, wright2014}. Previous studies have leveraged these similarities to predict and understand the weather and chemistry of WISE 0855's atmosphere \citep{skemer2016, morley2018, miles2020}. 

WISE 0855 is a field object in the solar neighborhood with an assumed age of 1-10 Gyr \citep{Luhman2014}. Evolutionary models for ages of 1-10 Gyr estimate the mass of WISE 0855 to be 3-10 $M_{jup}$ \citep{Luhman2024}, below the accepted deuterium burning limit of 13 $M_{jup}$, which is commonly used as the mass boundary between planets and brown dwarfs \citep{spiegel2011}. It remains uncertain whether WISE 0855 formed around a star and was subsequently ejected, or whether it originated as an unusually low-mass product of the star formation process. Regardless of its formation pathway, we classify any bound companions to WISE 0855 as exomoons, given its location within the planetary-mass regime following a precedent in the literature \citep{teachey, hwang2018,Limbach2021, sajadian2023, fu2025}.

Compared to planets orbiting bright host stars, observations of young FFPs that are still bright from the heat of formation can translate to higher-precision photometry that increases our sensitivity to smaller transits. Furthermore, since planets shrink as they evolve, these young FFPs can be larger in radius, which increases transit duration and transit probability (see Section \ref{1.3}). However, in the case where we can conduct sufficiently photometrically sensitive observations, older, colder objects also become viable locations to search for exomoons. This led \citet{Limbach2021} to highlight WISE 0855 as a compelling candidate for a transiting exomoon search, because despite its low temperature and luminosity, its proximity, in combination with the high sensitivity of JWST, make it possible to obtain high signal-to-noise ratio time-series observations. 

However, these low temperature conditions also favor cloud formation and disequilibrium chemistry. Specifically, these data were optimized to detect variability from water clouds forming at the top of WISE 0855's atmosphere. As a result, it becomes necessary to disentangle potential transit signals from surface variability features.

In this work, we search for transiting exomoons around WISE 0855 and use injection and recovery tests to demonstrate the sensitivity of JWST observations to Galilean moon mass analogs around FFPs. Section \ref{obsvandred} describes the details of the observations of WISE 0855 and our reduction of the data. We discuss our methods for constructing and fitting models to the light curves in Section \ref{sec3}, and report the results of our analysis and additional injection and recovery tests in Section \ref{resultssec}. In Section \ref{discussion}, we discuss our results and consider future opportunities to detect exomoons with JWST. Finally, we conclude by summarizing our findings in Section \ref{summary}.

\section{Observations and Reductions}
\label{obsvandred}
 
\subsection{Observations} 

The JWST NIRSpec observations of WISE 0855 were taken as a part of the Cycle 1 JWST GO program 2327 \citep{Skemerprop}.  WISE 0855 was observed at an RA of 08h 55m 3.51s and Dec of -07$^\circ$ 14\arcmin 33.47\arcsec on December 02, 2023 (UT) to obtain 11 hours of time-series spectra\footnote{The JWST data used in this paper can be found in MAST: \dataset[10.17909/rjtp-zn54]{https://doi.org/10.17909/rjtp-zn54}.}. Due to the object's high proper motion of ($\mu_{\alpha}$ cos $\delta, \mu_{\delta}$) = (-8.118 $\pm$ 0.008\arcsec yr$^{-1}$, 0.680 $\pm$ 0.007\arcsec yr$^{-1}$), and a parallax of 0.449$\pm$0.008\arcsec \citep{Luhman2016}, acquisition is non-trivial. For this reason, these observations adopted the coordinates, proper motion, and parallax values used by the GTO 1230 program that observed WISE 0855 just 7 months prior \citep{1230}. We acquired the target with wide aperture target acquisition (WATA) and observed in Bright Object Time-Series (BOTS) mode with a 1.6\arcsec $\times$1.6\arcsec square aperture to minimize flux losses that could potentially arise due to small misalignments of the target and a traditional slit. Based on recommended observing strategies and maximum signal-to-noise calculations, NRSRAPID readout and SUB2048 subarray, and the G395M/F290LP grating/filter were used. The data were collected during a single exposure broken into 44 15-min integrations with 996 groups per integration. The result is 44 R$\sim$1000 resolution spectra in the 2.87–5.27 $\mu$m wavelength range taken over a span of 11 hours. 
From the acquisition images, we confirm that the source appeared within the slit. Since WISE 0855 moved $\sim$4.9\arcsec in the $\sim$7 months since it was last observed, we examined the preceding GTO 1230 NIRCam images at the coordinates of our observations to confirm that no background galaxies were present to interfere with these data.

\subsection{Reductions}

\begin{figure*}
\includegraphics[width=\textwidth]{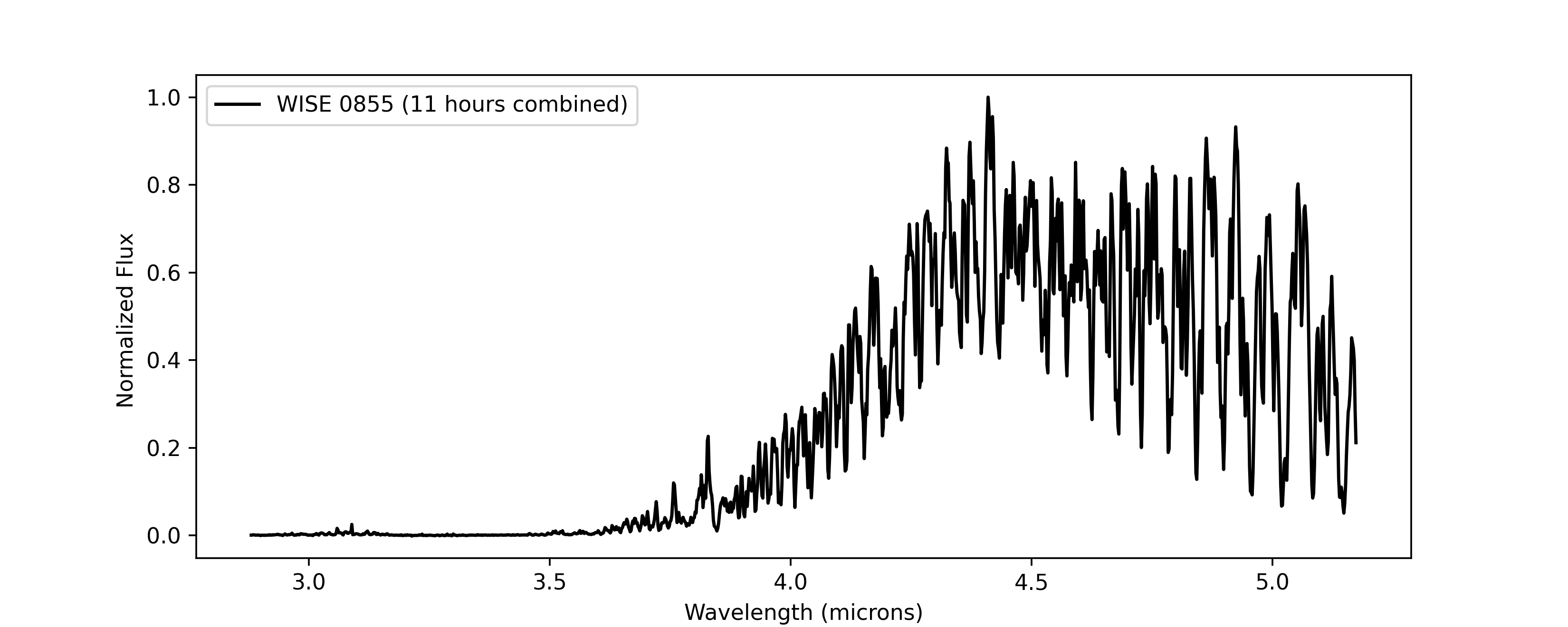}
\caption{Combined Spectrum of the 11-hour observation of WISE 0855 created from averaging the 44 spectra. The error bars are too small to be visible.}
\label{spectrum}
\end{figure*}

The JWST/NIRSpec time-series observations were reduced using version 1.14.0  of the standard JWST Pipeline \citep{bushouse_2024_12692459}. The pipeline was used with version `11.17.20' of the calibration reference data system (CRDS) and `\texttt{jwst\_1215.pmap}' context map. The reduction steps were published in \citet{rowland2024}, but we briefly summarize the process here. Stage 1 of the pipeline was used to turn detector ramps into slope images, then Stage 2 is used to produce flux calibrated spectral images. A custom extraction method is used on the reduced spectral images. We fit the curve of the spectral trace then extract over an 8-pixel wide aperture. The excess background is measured by using the median of the 3-pixels outside of the extraction aperture from both sides. 
The 44 extracted spectra are compared with each other to mask out hot pixels and other outliers. The results of the data reduction pipeline are shown in Figure \ref{spectrum} as a combined spectrum of the 44 15-minute integrations. To refine the data, we clip the shortest and longest wavelengths ends of the spectra to remove non-astrophysical flux information and interpolate over nans for a more robust mean.

\section{The Transit Search} \label{sec3}

\subsection{Exomoons around WISE 0855}
\label{1.3}
 
 Due to the faintness of WISE 0855, no high-resolution $v\,\sin i$ measurement exists to constrain a rotational inclination. In addition to this, despite the JWST light curve, there is still uncertainty in the rotation rate of WISE 0855 (discussed further in section 4.1). Therefore, to calculate transit probabilities of an exomoon, we must assume a random distribution of viewing angles. For the full range of possible inclination angles, we use the \citet{B1984} equation (eq \ref{probincl}), to determine the occultation probability for an exomoon of a given separation orbiting WISE 0855, where $R_{host}$ is the radius of WISE 0855, and $R_{moon}$ and $a$ are the radius and semi-major axis of an exomoon.

\begin{equation}
P_{occ} = (\frac{R_{host} + R_{moon}}{a})\approx\frac{R_{host}}{a} 
\label{probincl}
\end{equation}

\begin{figure}
\includegraphics[width=\columnwidth]{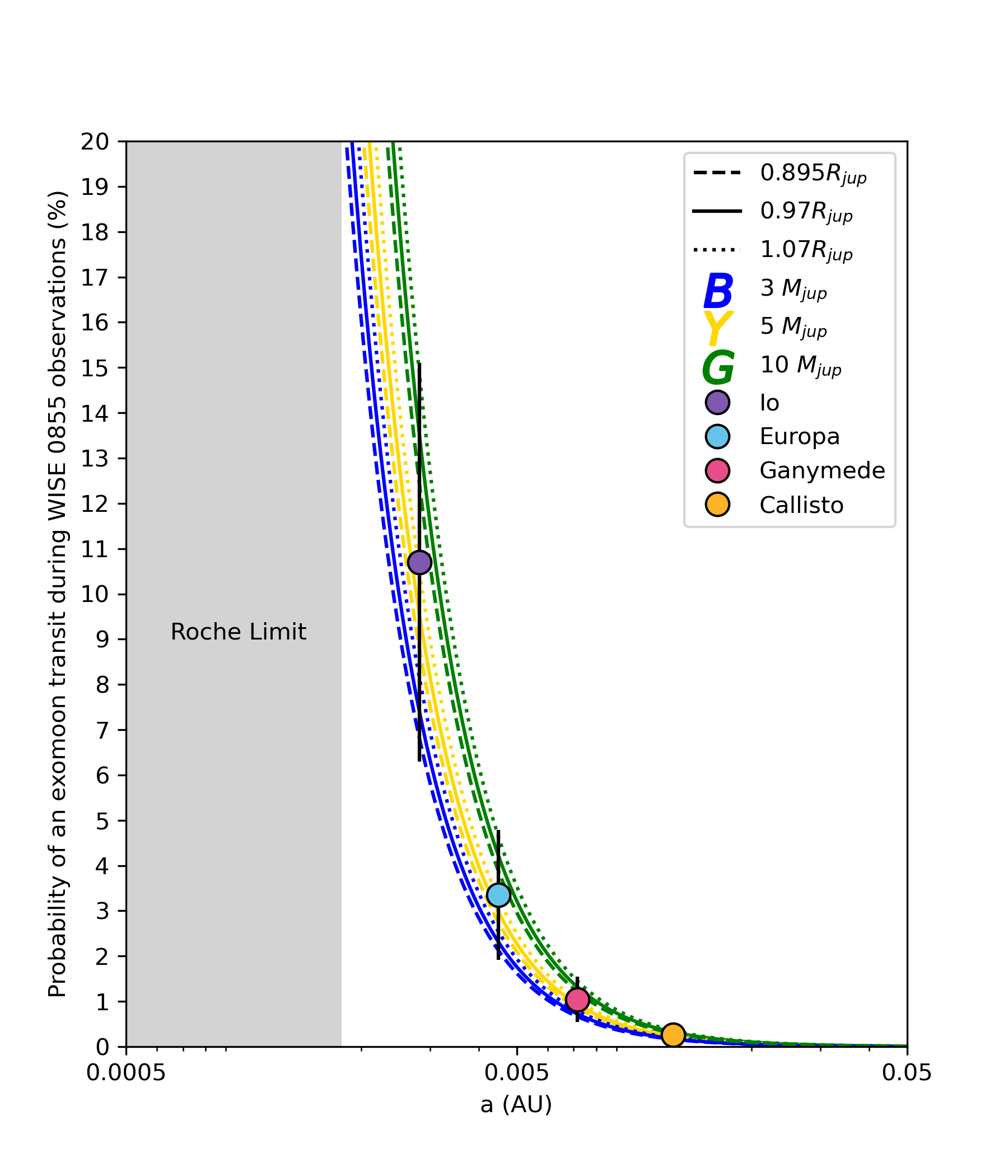}
\caption{Probability of an exomoon transit during the 11 hour WISE 0855 observation assuming a random orbital inclination angle. Different combinations of predicted radius and mass values for WISE 0855 are shown. The black line for each moon represents the extent of possible transit probabilities given these various mass and radius uncertainties for the host. The largest Roche limit for an Io-like moon around the densest prediction of WISE 0855 is at 0.00177 AU. Shown in gray is the distance range within which moons would be tidally disrupted.}
\label{prob}
\end{figure}

Additionally, for an edge-on inclination, the probability that an exomoon will transit during our 11 hour observation will depend on the exomoon's orbital period (eq. \ref{probobsv}).

\begin{equation}
P_{transit} = \frac{11 \text{ hours}}{T} = \frac{11 \text{ hours}}{2\pi\sqrt{\frac{a^{3}}{GM}}}
\label{probobsv}
\end{equation}
Here, $M$ is the mass of WISE 0855, and $T$ is the exomoon's orbital period in hours. Figure \ref{prob} combines these two probabilities by multiplying equation \ref{probincl} and equation \ref{probobsv} to show the overall probability that an exomoon would transit WISE 0855 during this 11 hour observation assuming a random inclination angle. The probability is shown as a function of separation from the host, with the separations of the Galilean moons as reference points. In calculating the probability, we include 3 different mass estimates that span the predicted 3-10 $M_{jup}$ mass range for WISE 0855 \citep{Luhman2014}. We also use 3 radius estimates: the maximum and minimum predicted radii values based on \citet{burrows2003} and \citet{saumon2008} evolutionary models for brown dwarfs at WISE 0855's luminosity (0.10-0.11 $R_{\odot}$), and the derived radius of 0.092 $R_{\odot}$ based on \citet{Luhman2024} JWST observations. 

We find that including the full range of possible mass and radii estimates is necessary because of the notable difference it makes in the transit probability. An exomoon at Io's separation orbiting a 3 $M_{jup}$, 0.092 $R_{\odot}$ WISE 0855 results in a 6.4\% transit probability while a 10 $M_{jup}$, 0.11 $R_{\odot}$ WISE 0855 results in a 15.1\% transit probability. Exomoons would be tidally disrupted inward of the Roche limit and cannot exist within this zone. This is shown in the gray region of Figure \ref{probobsv} with the largest Roche limit for an Io-like moon. 

As we consider transit probabilities, we also evaluate transit duration, calculated by eq \ref{bparameq} where $\Delta t$ is the transit duration of an exomoon around WISE 0855, and $i$ is the orbital inclination. 

\begin{equation}
\Delta t = \frac{TR_{host}}{\pi a}\sqrt   {\left( 1-\frac{R_{moon}}{R_{host}}\right) ^{2} - \left(\frac{a \ cosi}{R_{host}}\right)^{2}}
\label{bparameq}
\end{equation}


For edge-on systems, $i=90$, i.e the impact parameter $b=0$. The impact parameter is the sky-projected perpendicular distance between the path of the moon's transit and the center of the host \citep{Winn2014}. In such a case, and assuming $R_{moon} << R_{host}$, the equation becomes,

\begin{equation}
\Delta t \approx \frac{TR_{host}}{\pi a} \approx 2R_{host} \sqrt{\frac{a}{GM}}
\label{transitdureq}
\end{equation}

\begin{figure}
\includegraphics[width=\columnwidth]{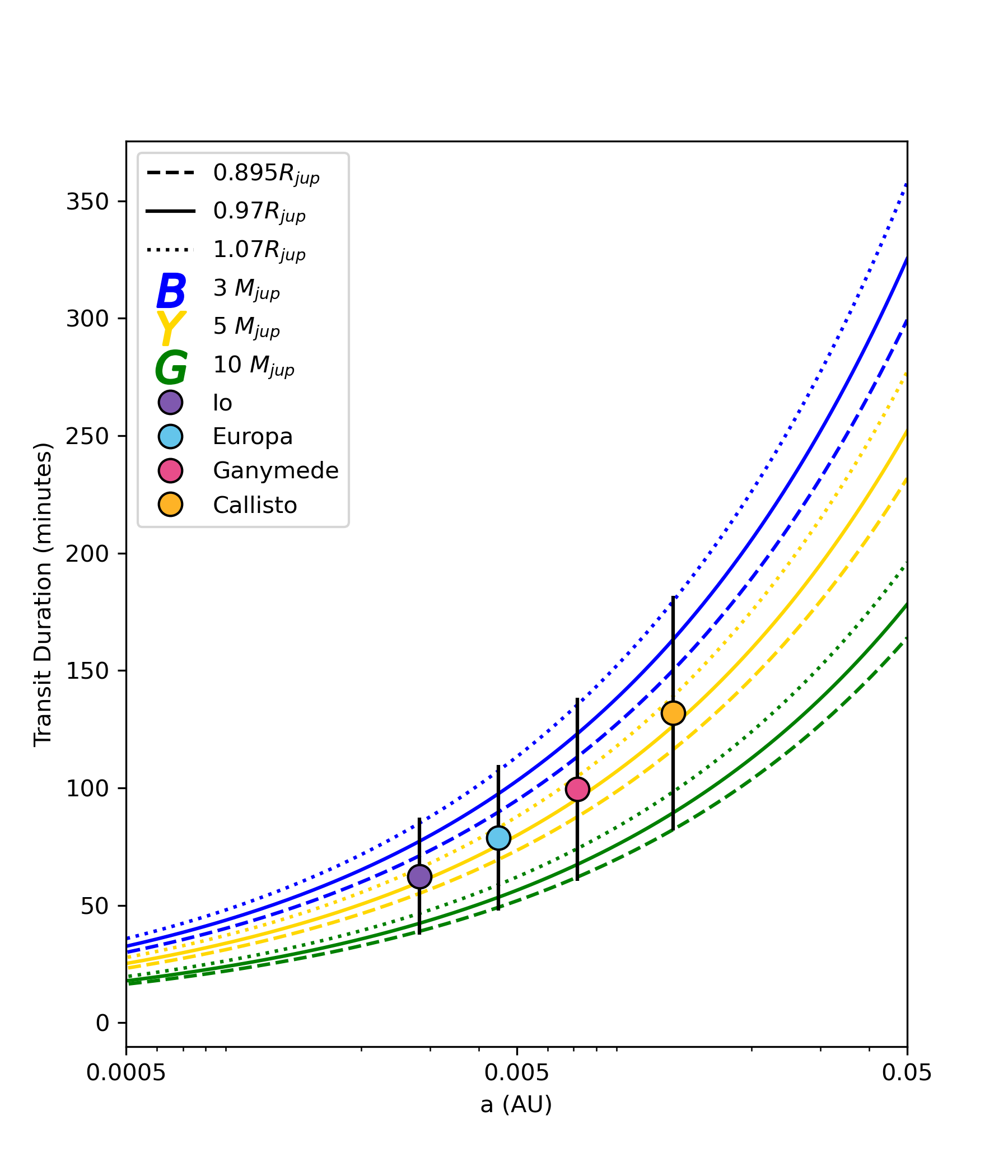}
\caption{Each curve represents transit duration for an edge-on system as a function of separation for a range of possible mass and radius values for WISE 0855 with the separations of the Galilean moons shown for reference. The black line emphasizes the extent of possible transit durations for each moon given these various radius and mass combinations.}
\label{transitdur}
\end{figure}

Figure \ref{transitdur} illustrates the increase in transit duration for a moon at increasing separation assuming an edge-on system. For reference, the separations of the Galilean moons are shown, where an exomoon at an Io-like separation would transit WISE 0855 for a duration of 62.4 min. 

\subsection{WISE 0855 Light Curves}
Clouds and chemical processes in the atmospheres of substellar objects can cause wavelength-dependent variability in the observed flux \citep{biller2024, mccarthy2025}. In contrast, true transits are expected to be gray (to first order), producing a consistent transit depth across all wavelengths. To distinguish between host variability and true transit signals, we can construct two light curves from spectral regions that exhibit different variability behavior. While it is possible that the host's intrinsic variability features may mimic transit-like dips in one light curve of a particular wavelength, a second light curve at another wavelength often results in a different depth. This provides a means of ruling out false positives attributed to variability \citep{biller2024, mccarthy2025}. In this work, we opt to use two light curves from the WISE 0855 dataset to search for any identical and concurrent transit signals that may be indicative of an exomoon transit.

\begin{figure*}
\includegraphics[width=\textwidth]{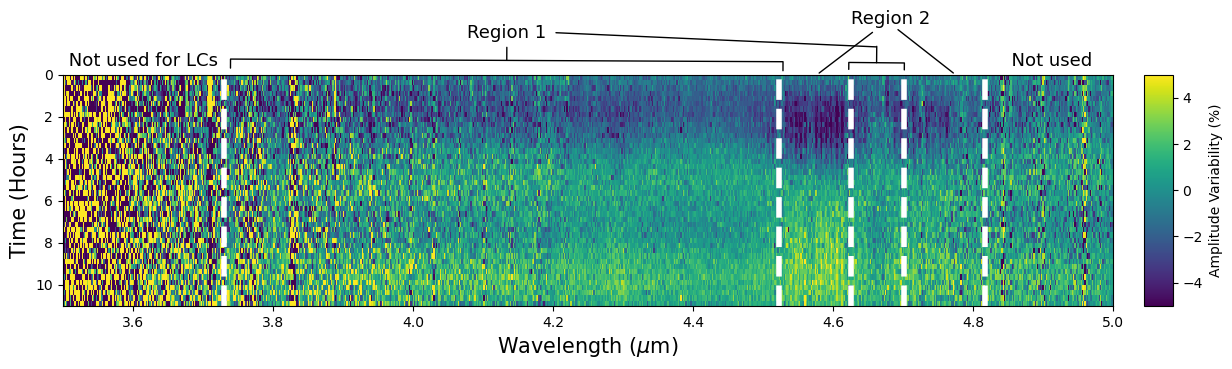}
\caption{The time-series spectra with the full wavelength range (in microns) of the observations shown on the x-axis, observation time (in hours) on the y-axis, and a color bar representing the amplitude variability for any given flux value as a percent deviation from the mean. The 2 wavelength regions used to construct the light curves are shown where each has a distinct variability pattern. Region 1: negative, positive, negative positive amplitude; Region 2: negative, positive amplitude. The picket fence pattern (narrow striping) visible in region 2 is due to CO absorption.}
\label{2d}
\end{figure*}

Rather than using evenly spaced wavelength bins, we separate the data into two regions with distinct variability patterns to maximize their differences. These features are illustrated by the variability map in Figure \ref{2d}, where the data are described by their deviation in amplitude from the mean. Region 1 spans 3.73-4.52 $\mu$m, and 4.62-4.70 $\mu$m,  with a dark, bright, dark, bright pattern over time while Region 2 spans 4.52-4.62 $\mu$m and 4.70-4.82 $\mu$m with dark, then bright features that roughly corresponds to the CO band head. 
We choose to omit the 2.87-3.74 $\mu$m range of the spectra that features lower signal-to-noise data (see Figure \ref{spectrum}). The lower flux at this bluer wavelength range is a consequence of the low temperature of WISE 0855. Furthermore, while the reddest region, spanning 4.82-5.27 $\mu$m, matches the variability pattern of region 1, different pipeline reductions (versions 1.12.5 and 1.14.0 of the standard JWST Pipeline) resulted in inconsistent variability patterns. To ensure a more conservative approach in our data analysis, we omit this region as well.

\begin{figure}
\includegraphics[width=\columnwidth]{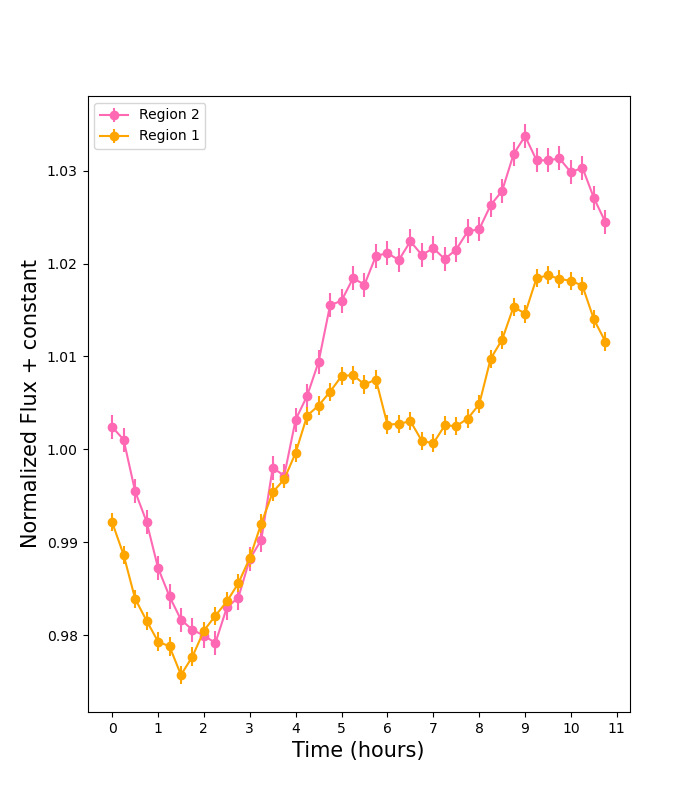}
\caption{The light curves, normalized to 1, of regions 1 and 2 built from a weighted average calculated using empirical errors.}
\label{lcs}
\end{figure}

For each region, we build a corresponding light curve from a weighted average of the spectra at every time step. We empirically estimate the error at every wavelength for use in the weighted average by computing the standard deviation of the differences between flux values at adjacent time steps and dividing by $\sqrt{2}$ (standard error propagation). By fitting the resulting weighted average light curve of each region with a 7th degree polynomial, we can subtract the overall shape of the host variability to then repeat the error propagation on the flattened data, allowing us to calculate an overall standard error of the weighted average for each light curve that reflects the photometric uncertainty, independent of the host’s variability.  We calculate errors of 0.098\% and 0.13\% for regions 1 and 2, respectively. Figure \ref{lcs} shows these weighted average light curves, normalized to 1, for each region.


\subsection{Light Curve \& Transit Fitting Models}
To investigate the statistical probability of the presence of a transit in our data, we employ a Gaussian Process (GP) model with a Quasi-periodic kernel based on \citet{van2018} and \citet{limbach2024} models. The GP model is non-parametric and returns a Gaussian distribution of functions that describe the host's variability in flux as a function of time. We compare the likelihoods of two models: a GP + transit model and a GP only model. The GP + transit model is fit simultaneously to both light curves and requires that the transit model is identical in both. The transit is fit as a trapezoid with transit mid-time, depth, duration, and slope as free parameters. We do not take limb darkening into account as the temporal resolution of our data does not warrant that complex of a model. The GP only model does not include a transit model, meaning any variability in flux is assumed to be intrinsic to variability in the atmosphere of WISE 0855. This captures the unique variability signals in each light curve separately.

Under the assumption of a ``gray," transit feature, we simultaneously fit the two light curves with our GP + transit model, requiring that the presence of a transit in one light curve be concurrent with a corresponding feature in the other light curve. By simultaneously fitting multiple light curves, we improve our constraints on which dips in flux will be identified as transits, limiting the extent to which the host variability can mimic these features.

We select our model priors based on the predicted characteristics of potential companions around WISE 0855 presuming that our solar system's gas giants host moons representative of the greater exomoon population \citep{Limbach2021}.
As illustrated in Figure \ref{transitdur}, we can compute the expected transit durations for Galilean-like exomoons around WISE 0855, with an exomoon at an Io-separation being the most likely to transit (Figure \ref{prob}). Since the inclination of WISE 0855 is unknown, we adopt a $\sim$1 hour upper limit of expected transit durations at this separation, where instances of an impact parameter b$\neq$0 would decrease this value further. Based on the increased transit probability for closer-in companions, and the upper limit for an exomoon at an Io-separation, we select a uniform transit duration prior for exomoons at all separations to be 0-80 min with the GP + transit model and search for maximum transit depths of up to 1.3\%. We allow all possible impact parameters and for the transit mid-time to be at any point in the light curve, opening up the possibility to identify grazing transits as well.

We run the GP + transit model and the GP only model on the 2 light curves.  Using \texttt{dynesty} \citep{speagle2020,koposov2022}, a nested sampling package, we determine the Bayesian evidence for each model, represented by a log$z$ value. For the GP + transit model fit we use 400 live points with an uncertainty in the \texttt{dlogz} of 0.8, while the GP only model uses 10 instances of 60 live points with an uncertainty in the \texttt{dlogz} of 2.  In both models we use the default bound. The resulting effective \texttt{dlogz} uncertainty for the GP only models is $\sim$0.7 which is equivalent to doing a single run with this precision, but with a faster run-time efficiency since a single run of this efficiency can get stuck or fail to converge. The more favored model of the two will have a higher log$z$ value, where the $\Delta$log$z$ between them corresponds to the strength in evidence shown in Table \ref{table:1}.

\begin{center}
\begin{table}
\begin{tabular}{ | c |c| } 
 \hline
 $\Delta$log$z_{final}$ value & Strength of Evidence  \\ 
 \hline
 0 to 1.5 & Not worth mention  \\ 
 \hline
 1.5 to 2.3 & Substantial \\ 
 \hline
 2.3 to 4.6 & Strong  \\ 
 \hline
 $>$ 4.6 & Decisive  \\ 
 \hline
\end{tabular}
\caption{$\Delta$log$z$ Strengths of Evidence}
\label{table:1}
\end{table}
\end{center}

The GP + transit model requires that the two light curves be fit simultaneously, producing one combined log$z$ value. The GP only model fits the light curves independently, computing two individual log$z$ values, each with its own error. To compare the Bayesian evidence of the two models, we sum the two individual GP only model values for a combined log$z$ and add the errors in quadrature.  

Since it is possible that the GP + transit model could be favored over the GP only model due to features in the host variability that mimic real transits, we calibrate out this bias by correcting the resulting $\Delta$log$z$ value. To calculate a correction factor, we run both models again on flipped versions of the same light curves. When we flip the light curves, no transits will be present, as they can only reduce flux. However, assuming the variations in flux are close to symmetric about the mean with comparable increasing and decreasing features, we're left with light curves that are still representative of the host variability. Any preference for the GP + transit model over the GP only model will give us a $\Delta$log$z$ representing the extent to which features of WISE 0855's variability emulate transit signals within our light curves. We require that any resulting $\Delta$log$z$ values we consider from the unflipped models must be above this threshold. By subtracting the flipped model's $\Delta$log$z$ from the unflipped, we calculate our  $\Delta$log$z_{final}$ value from which we determine the probability of an exomoon transit detection based on the strength of evidence.



\begin{figure*}
\begin{centering}
\includegraphics[width=14cm]{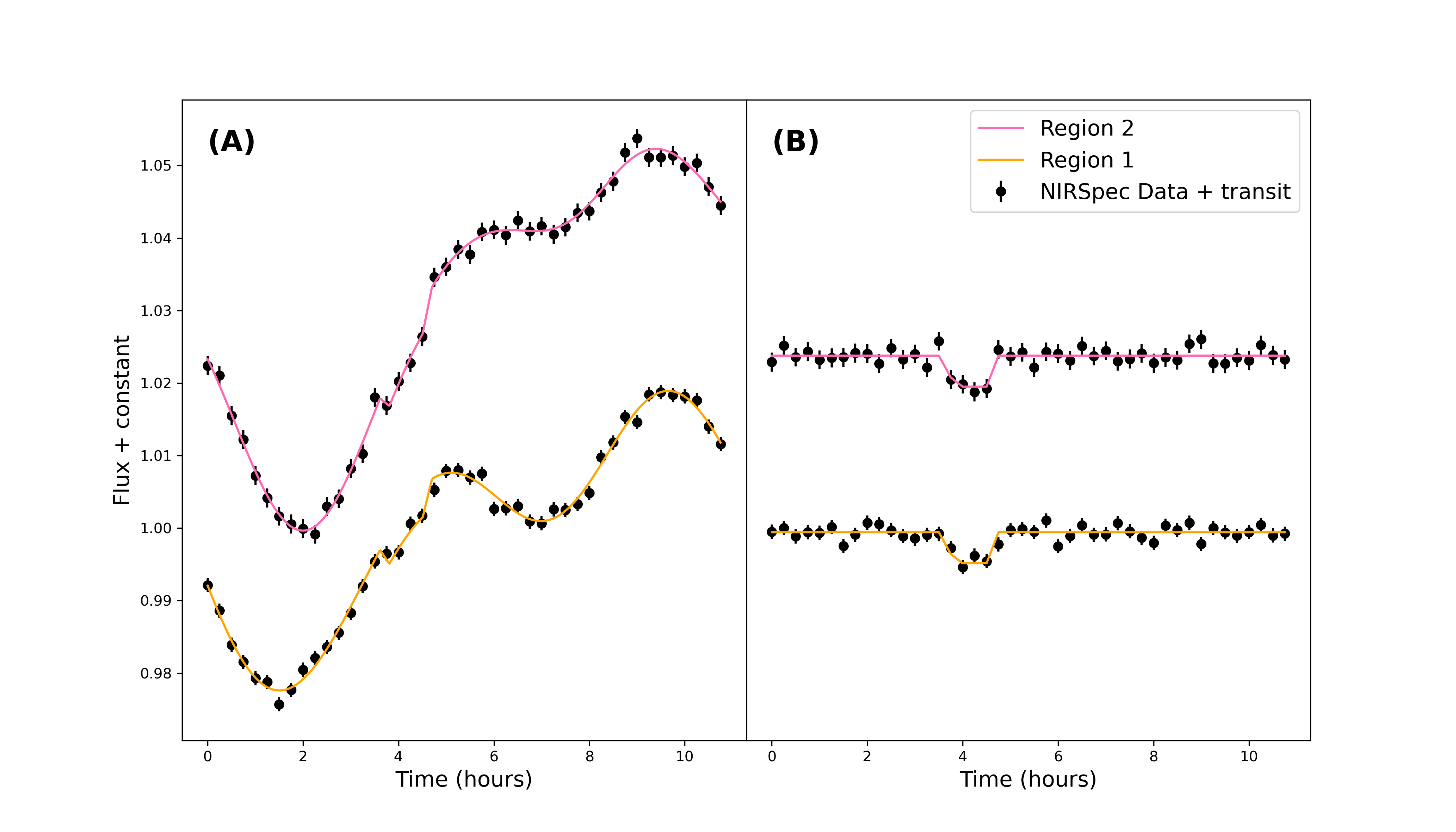}
\caption{Panel A: The light curves of WISE 0855 built from the two different wavelength regions are shown in black with an injected 0.3\% transit (Mars-sized) at $\sim$4.2 hours. The solid lines represent the fit produced by the GP + transit model.; Panel B: The flattened light curves of each region are shown after subtracting the GP portion of the GP + transit model's fit, revealing the injected Mars-sized 0.3\% transit at $\sim$4.2 hours.}
\label{exinj}
\end{centering}
\end{figure*}
\section{Results} \label{resultssec}
\subsection{Transit Search Results}
For the unflipped case, we calculate a GP + transit model log$z$ of 486.98$\pm$0.45 and a GP only model log$z$ of 475.58$\pm$1.35, resulting in a $\Delta$log$z$ of 11.41$\pm$1.43 in favor of the transit model. The flipped light curve models result in a log$z$ of 486.73$\pm$0.43 for the GP + transit model and 475.76$\pm$1.35 for the GP only model with a $\Delta$ log$z$ of 10.97$\pm$1.42 in favor of the transit model. This flipped $\Delta$log$z$ provides us with a correction factor we can subtract from the unflipped light curve's $\Delta$log$z$ to calculate  our $\Delta$log$z_{final}$ value of 0.43 in favor of the GP + transit model. Our strength of evidence falls within the \emph{Not worth mention} range.  To confirm these results and to demonstrate the validity of the chosen regions, we repeat this procedure with a +25nm and +50nm shift in the chosen wavelength ranges and do not find a significant difference in the resulting $\Delta$log$z_{final}$ value.  

Although the GP + transit model results sufficiently demonstrate that the identified transit signature is not statistically significant, we further assess its plausibility by examining the properties of the recovered transit as it appears in the original (unflipped) dataset. The recovered transit depth of 0.22\% is consistent with a $\sim$0.53 $R_{\oplus}$ sized moon. The peaks of the posterior distribution for the recovered transit mid-time align with the large dips in flux we see in the light curves (see Figure \ref{noinjgpt} and Figure \ref{reggpt} in the appendix). Furthermore, the shape of the transit is consistent with a grazing event, and, with a duration of $\sim$2 hours, represents a moon that is well-separated from the host. As illustrated in Figure \ref{prob}, we are far less likely to observe moons at far separations compared to close-in moons. Therefore, in addition to the lack of statistical significance, this large transit feature is likely attributable to the host's variability. There are regions of the light curve where the variability does not differ drastically between the two regions. As a result, it becomes more difficult to detect transits in these regions because we are variability limited. We conclude that there is no evidence for a transit detection supported by the data. 




 The results of our models may also provide an opportunity to begin to understand the variability features of the host due to atmospheric conditions. This data set only represents the second long-duration time-series observations conducted for WISE 0855. The Spitzer data, \citet{esplin2016} suggests the potential rotational periods for WISE 0855 of $\sim$10.8 and $\sim$13.3 based on the two epochs of their [4.5] light curves, with inconclusive results for a single rotational period due to the irregularity of the variability. Our models also report a rotational period for the host with region 1 of the GP + transit model returning 14.35 $^{+10.32}_{-8.23}$ hours, and region 2 returning 8.57 $^{+6.24}_{-5.64}$ hours. Our GP only models suggests rotation periods longer than 14 hours for both regions. We find that longer duration time-series observations of WISE 0855 are needed to draw definitive conclusions about the rotational period. However, based on the temperature and spectrum, it is clear the low temperatures and atmospheric features are having significant effects on the variability signatures we see in the data.   


\subsection{Injection and Recovery Tests }\label{injandretsection}
While we do not find an exomoon in these observations, we use injection and recovery tests to identify the parameter space in which transits would have been detectable if they had occurred during this observation. We perform 50 random injections for 6 transit depths: 1\%, 0.5\%, 0.4\%, 0.3\%, 0.2\%, and 0.1\%. The chosen depths cover a range of exomoon sizes that are theoretically predicted to be commonly occurring around WISE 0855, but more specifically, a 1\% transit represents an Earth sized exomoon, 0.2\% represents a Ganymede sized exomoon, and perhaps most interesting, a 0.3\% transit represents a Mars-sized exomoon whose mass ratio to WISE 0855 is akin to the mass ratio between Io and Jupiter. As shown in Figure \ref{prob}, closer-in moons, like Io, correspond to higher transit probabilities.

To demonstrate the performance of the GP models and our ability to fit and remove host variability, we highlight an example of an artificial 50 min transit of 0.3\% depth randomly injected at 4.28 hours in panel (A) of Figure \ref{exinj}. The solid lines represent the GP + transit model's best fit for each light curve. We are able to subtract the variability with the GP, leaving the transit and a flattened light curve as shown in panel (B). The recovered mid-time of the transit reported by the model is 4.20 hours, a result within 2 minutes of the injected time. For this example, the GP + transit model is favored with a $\Delta$log$z_{final}$ of 7.69, corresponding to \textit{Decisive} confidence\footnote{See the appendix for the corner plots for this example and the relevant models described in 4.1.}. This highlights the success of the GP models to identify the transit feature as being distinct and separate from WISE 0855's characteristic variability.



\begin{figure*}
\begin{centering}
\includegraphics[width=17cm]{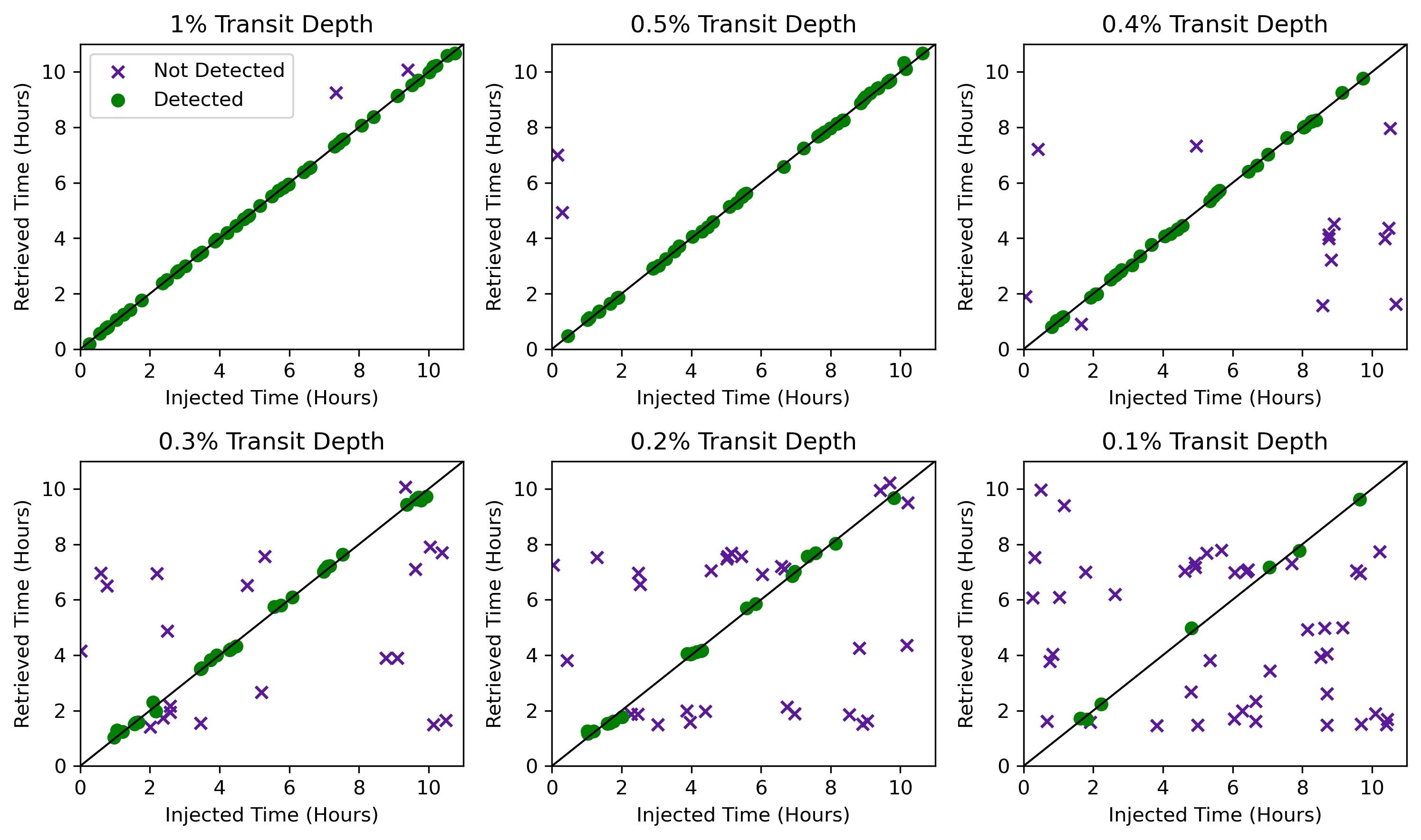}
\caption{The results of the 50 injection tests for each of the 6 transit depths where the green points are detections i.e. the retrieved time is within 15 minutes of the injected time, and purple Xs are non-detections.}
\label{6panelrets}
\end{centering}
\end{figure*}

For each of the artificial transit injections, we follow the same log$z$ comparison process from the no-injection case, again using the empirical $\Delta$log$z$ correction factor of 10.97, to calculate a $\Delta$log$z_{final}$ for each injection. We consider a successful detection to be one whose recovered mid-time of transit is within 15 minutes of the injected transit time. In Figure \ref{6panelrets} we show the results of each test with the transit mid-time retrieved by the GP + transit model as a function of the injected time. The clumping nature of successful detections, seen clearly in the 0.3\% and 0.2\% panels, is a result of transits being easier to detect in some portions of the light curve. This indicates that in the regions of the light curve where there are neither detections nor non-detections along the diagonal, we are likely limited in our ability to detect transits due to the host's variability.

\begin{figure*}
\begin{centering}
\includegraphics[width=17cm]{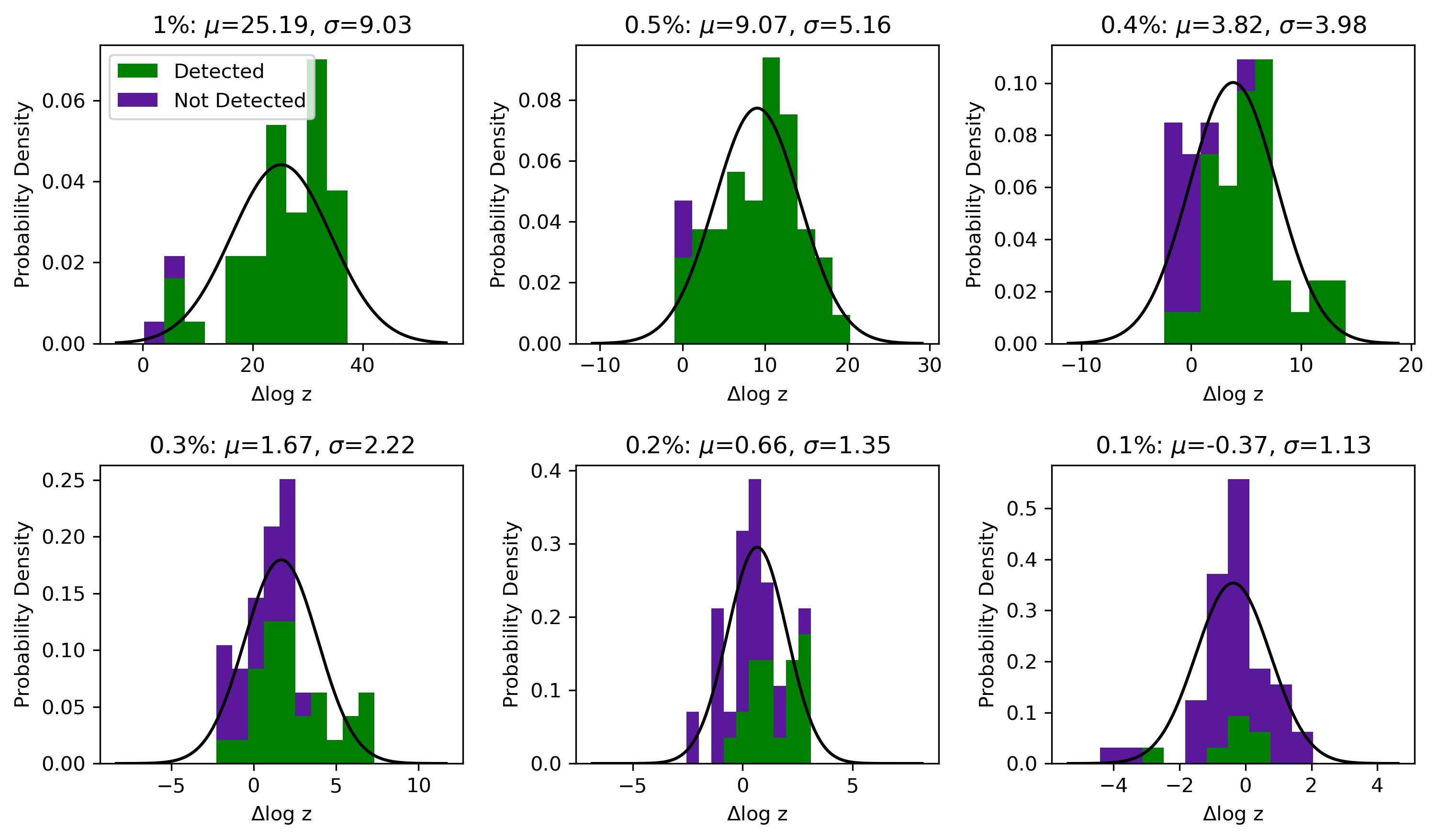}
\caption{Histograms of the 50 $\Delta$log$z$ values for each transit depth with non-detections in purple stacked on top of the successful detections, i.e. retrieved transit time within 15 min of injected time, in green.}
\label{6panelhist}
\end{centering}
\end{figure*}

In a blind transit search, we must rely on the strength of evidence as described by the $\Delta$logz values derived by the models to consider the validity of a transit signature rather than the mid-time match. In Figure \ref{6panelhist} we again show the results of the injection and recovery tests, but with a focus on each transit depth's $\Delta$logz distribution. The number of detections increases with transit depth, as reflected in the distributions. However, while there is a strong correlation between the $\Delta$logz values and mid-time detections, it is not quite one-to-one. For instances where the $\Delta$logz is high, but there was no mid-time match (a match is constituted by a retrieved mid-time within 15 minutes of the injected time), this would be seen as a false positive detection in a blind transit search. However, there is only 1 instance of false-positive for a $\Delta$logz value corresponding to \textit{Decisive} confidence out of the 300 total injection tests.

\begin{figure}
\includegraphics[width=\columnwidth]{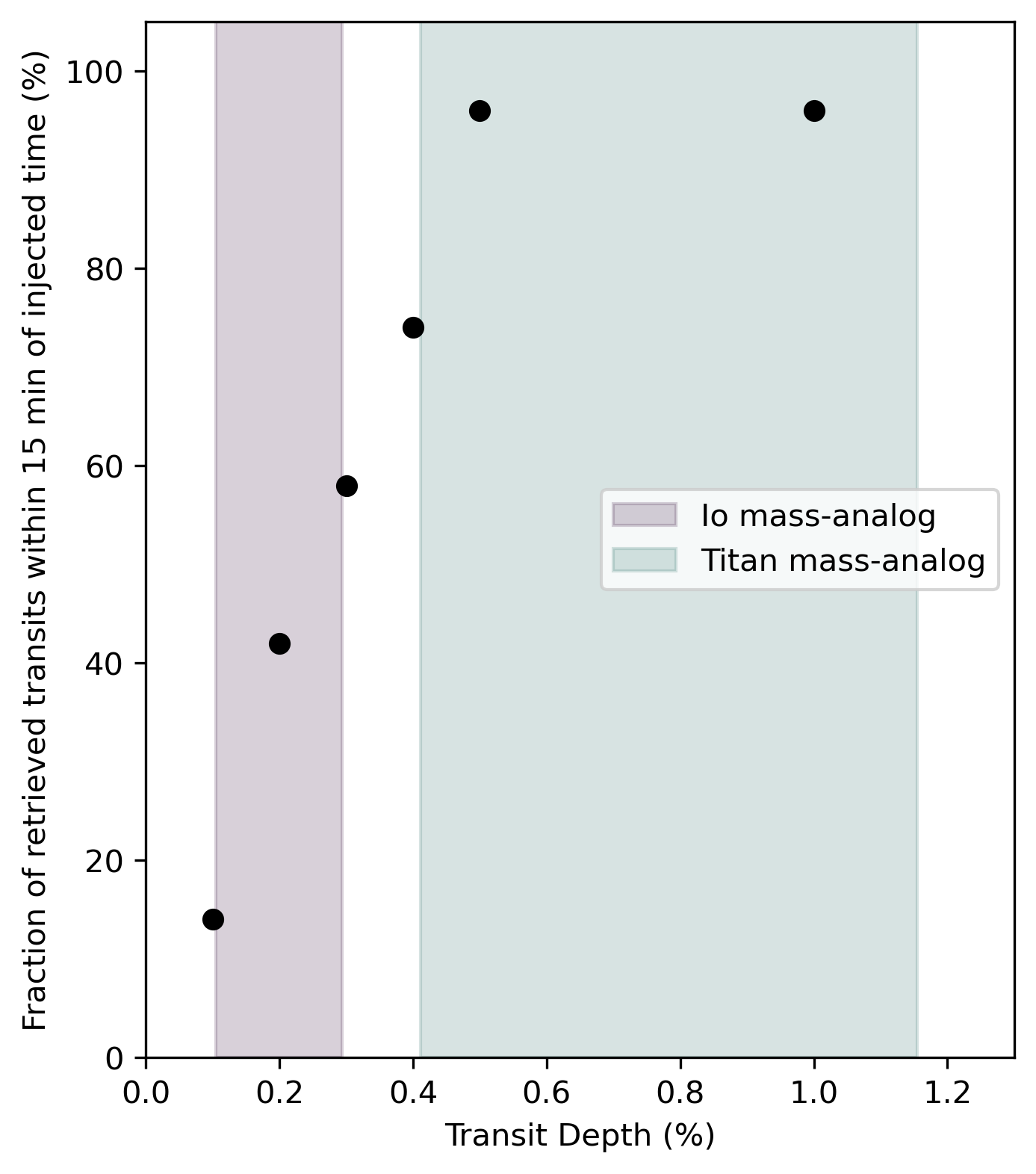}
\caption{The fraction of 50 transit injections recovered by the GP + transit model within 15 minutes of the injected time for each transit depth. The shaded regions represent the range of expected transit depths for Io and Titan mass analogs based on the predicted mass and radii values of WISE 0855.}
\label{finalfraction}
\end{figure}

To summarize these results, Figure \ref{finalfraction} shows the fraction of successful detections for each transit depth. We detect 96\% of the 1\% and 0.5\% transit depths, and 74\%, 56\%, 42\%, and 14\% of the 0.4\%, 0.3\%, 0.2\%, and 0.1\% transit depths respectively. Included as purple and blue shaded regions are the transit depths of the corresponding exomoon mass analogs for Io and Titan. We choose to include the following mass analog analysis to ground our results in satellite formation theory. Transiting exomoons make radius measurements our observable, but mass may be the more fundamental parameter to consider \citep{canup2006}. By focusing on host size alone, we overlook the significant influence that the predicted mass of WISE 0855 may have on potential companions that would subsequently be more massive, and therefore larger, than Io and Titan analogs derived only from WISE 0855's radius.

If we assume a total moon mass to host planet mass ratio of $\sim10^{-4}$ as proposed by \citet{canup2006}, then based on the predicted mass range of 3-10 $M_{jup}$ for WISE 0855, the system would have a total moon mass between 0.10-0.32 $M_{\oplus}$. It follows that individual Galilean-like or Titan-like exomoons would also scale with their host mass. To estimate potential Titan-like and Io-like exomoon masses around WISE 0855, we apply the mass ratios of both Io-to-Jupiter and Titan-to-Saturn to the predicted mass range of 3-10 $M_{jup}$ to calculate Io and Titan mass analogs. With these mass ranges, we determine the expected corresponding radius values with the \citet{Chen2017} relationship of $R\sim M^{0.28}$ for Terran worlds. The radii of these mass analogs result in the transit depth ranges shown in Figure \ref{finalfraction} that will depend on the predicted size of WISE 0855.  
The lower limit of each range corresponds to the smallest predicted exomoon mass analog transiting the largest predicted size of WISE 0855 while the upper limit corresponds to the largest predicted exomoon mass analog transiting the smallest predicted size of WISE 0855. The results of our analyses show we are sensitive down to Titan mass analogs and possibly Io mass analogs in these observations.

\section{Discussion} \label{discussion}

The high precision of JWST makes it the most efficient existing facility for conducting exomoon searches using this method. Although our target is faint, the telescope's sensitivity in combination with the absence of a nearby host star, allows us to reach down to the photon noise limit in some parts of the light curve. With an error of 0.098\% in the region 1 light curve and 0.13\% in the region 2 light curve per 15 minute exposure, the 5$\sigma$ detection threshold for a 1 hour transit is 0.245\% (1470 ppm), and 0.325\% (1950 ppm), respectively.  In other portions of the light curve, notably where we missed detections at transit depths $\geq$0.2, we were likely variability limited, rather than photon noise limited. However, a natural consequence of the fact that these observations were designed with WISE 0855's clouds as the primary science case is that our target is significantly variable. By targeting the least variable FFPs, we open the opportunity to reach down to more sensitive regimes. 

Although less variable signals could be indicative of less atmospheric activity, they may also be related to the inclination angle of the host \citep{vos2017}. Regular satellites are expected to form in the equatorial plane of their host, but these orbits may change over time due to dynamical interactions. For example, \citet{Rabago2019} show that for ejected planets, which have a $\sim$85\% chance of retaining close-in moons on Galilean orbits, the inclinations of the satellites' orbits may be disrupted. In future studies, it is therefore reasonable to investigate the performance of our GP models even for systems that are not edge-on. Relatedly, the cycle 4 JWST GO program 8846 \citep{nguyenprop} will observe 3 L-T transition brown dwarfs and planetary mass objects with inclinations varying from pole-on to edge-on to better understand the correlation between color, inclination, and host variability. Such a data set can provide the opportunity to understand how our model's transit recovery performance varies with host inclination. 

Designing observations to optimize exomoon detections (rather than with studying weather as the primary science case) could further improve our results. Longer duration observations would open up the opportunity for catching exomoons on longer orbital periods or for observing multiple transits to aid in detection confirmation. Furthermore, higher cadence observations will provide the models with more light curve information to allow for the distinguishing between variability features and transit signatures. This refinement in future methodology is closely tied to the potential to further develop the GP models. Observations that cover a longer wavelength range, such as using NIRSpec PRISM \citep{mccarthy2025}, could give rise to more than two distinct wavelength regions for the models to compare. While WISE 0855 is faint, it's proximity enables sufficient conditions for a transit search. In most cases, however, it will be by targeting young FFPs that are still bright from the heat of formation that will allow for high signal-to-noise observations \citep{Limbach2021}. Furthermore, understanding the variability properties of even the most variable substellar objects, like VHS 1256b \citep{Miles2023}, will improve our ability to identify exomoon and exosatellite signatures for FFPs and bound wide-orbit exoplanets overall. The method we describe in this paper is not exclusive to FFPs, but can be used to search for moons in the light curves of wide-orbit exoplanets and some directly imaged exoplanets with NIRCam's dual-band imaging \citep{zhangprop, brandnerprop, ruffioprop, mullensprop}.

\begin{figure}
\includegraphics[width=\columnwidth]{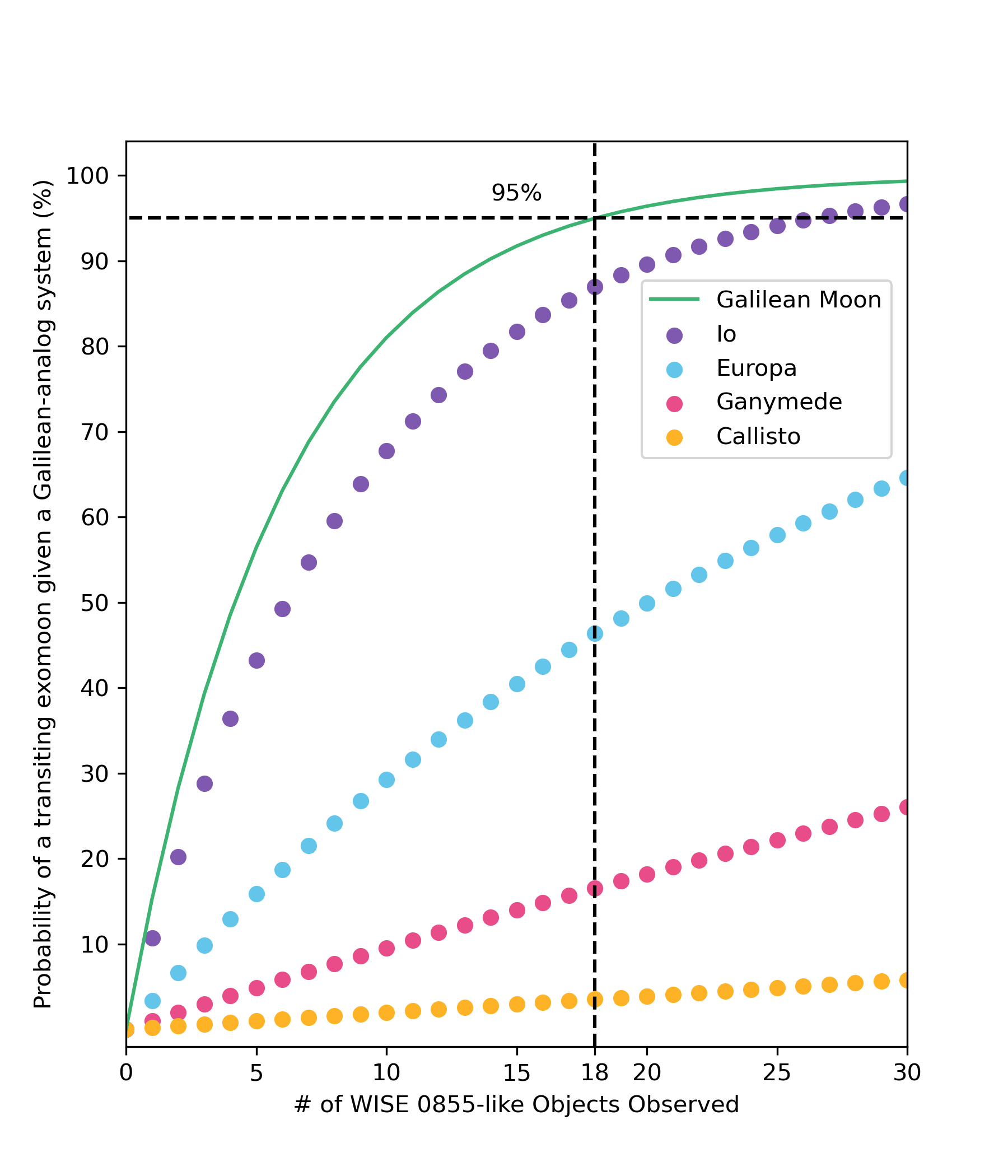}
\caption{Probability of a transiting exomoon around WISE 0855 assuming a mass of 6.5 $M_{jup}$ and a radius of $\sim$1 $R_{jup}$, random inclination angle, and 11 hour observation. Based off these conditions, we calculate the probability of a transit for exomoons at each of the Galilean moon separations for an increasing number of objects. The combined probability results in a 95\% chance of a transiting exomoon after observing 18 WISE 0855-like objects that have Galilean analogs.}
\label{numobsvs}
\end{figure}

Figure \ref{prob} explores the probabilities of detecting an exomoon around WISE 0855 at separations similar to those of the Galilean moons. Being that the inclination of WISE 0855 is unknown, and that these observations span 11 hours, the probability that we would have observed a transiting moon at an Io separation spans 6.3-15.1\% depending on the true mass and radius value of our host. This represents the probability for a single dataset, but Figure \ref{numobsvs} illustrates how the likelihood would improve with more observations of WISE 0855-like objects. Under the same observational conditions, i.e. 11 hour observations of hosts with unknown inclination angles, an average mass of 6.5 $M_{jup}$ and a radius of $\sim$1 $R_{jup}$, we can consider each target to have a 10.7\% chance of hosting an Io, a 3.4\% of hosting a Europa, and a 1\% and 0.5\% of hosting a Ganymede or Callisto, respectively (see Figure \ref{prob}).  If all of the systems have Galilean moon analogs, then with each observation, our likelihood of catching a transit of any one of them culminates in a 95\% chance after 18 observations.

The presence of the transit signal in our data, however, does not guarantee its identification by our models. Based on our injection/recovery tests, the probability of a successful detection will vary depending on the transit depth of said exomoon and the underlying host variability. For example, with a detection rate of 96\% for depths $\geq$0.5 (see Figure \ref{finalfraction}), this suggests that there would be a 95\% chance that an exomoon transit is present in our dataset, and a 96\% chance that our models will recover it. Therefore, we have a $\sim$91\% chance overall of detecting an exomoon with a transit depth $\geq$ 0.5 after 18 such observations if the targets have moons in a Galilean orbital configuration. As shown in Figure \ref{finalfraction}, this transit depth is reflective of Titan mass-analog exomoons.

The probabilities taken into account for Figure \ref{numobsvs} consider the specific observational conditions of this dataset, but the overall probability of exomoon transit recovery will continue to improve by targeting systems with known inclinations that are close to edge-on assuming that moons are more preferentially aligned with the spin-axis of their host. Furthermore, we can derive a more precise probability for hosts with better constrained mass and radius measurements, such as objects with measured dynamical masses or objects that are confirmed members of young moving groups. 

The popularity of brown dwarf variability science in recent JWST cycle proposals is conducive for transiting companion searches \citep{2965, 3181, 3548, 3496, 3375, 4758, 5226, 6139, 6474, Limbachprop, nguyenprop}. In particular, cycle 4 JWST GO program 8155 \citep{Limbachprop} will observe $\sim$50 substellar objects between 2 and 75 $M_{jup}$ with dual-band NIRCam imaging. Although many of the targets are above planetary mass, this data will contribute to the overall development of exosatellite detection techniques and, in a single program, cover the number of objects we predict are necessary to catch a satellite as it transits in front of its host.

If theoretical predictions are correct and if solar system analog moons are common, then the sample of substellar worlds JWST will monitor is large enough that we would expect to detect multiple transiting companions. If no detections are made this will provide strong evidence that the occurrence rate of exomoons is different than our solar system moons. Regardless of if a transit is detected, this program, along with the dozens of JWST programs that have been selected to monitor other exoplanets, FFPs, and BDs, will be useful for optimizing data reduction techniques, model development, and conducting more artificial injection tests in future exomoon studies.



\section{Summary}\label{summary}

We have carried out the first search for exomoons around a planetary mass object with JWST. We analyze 11 hours of JWST NIRSpec observations of WISE 0855 in search of transiting exomoon signatures. WISE 0855 is an ideal target due to its proximity and free-floating nature.  As a planetary mass Y dwarf, we expect any potential satellites could be representative of the exomoon population. We fit two models, one that assumes an underlying transit and one that does not, to light curves built from distinct wavelength ranges. Our GP+Transit model requires the presence of a concurrent, identical transit signature in both light curves, under the assumption the transit feature is wavelength-independent. After fitting our models, we find no substantial evidence for a transit. 

Injection and recovery tests of artificial transits result in detection rates of 96\% for 1\% and 0.5\% depths, and 74\%, 56\%, 42\%, and 14\% of the 0.4\%, 0.3\%, 0.2\%, and 0.1\% transit depths, respectively, where a successful detection is a recovery time within 15 minutes of the injection. Based on the probability of a transiting exomoon around WISE 0855, we calculate a 95\% chance of observing a Galilean analog exomoon after 18 similar observations, assuming they each host Galilean-like systems (Figure \ref{numobsvs}). The probability of the successful detection of said exomoon will then depend on the transit depth (Figure \ref{finalfraction}). These odds will only improve as we design future exomoon-focused observations, improve the precision of our GP models, and target objects with known inclinations or more than 4 major moons.

We have demonstrated, for the first time, that JWST is capable of detecting moons analogous in mass-ratio to those in the Solar System. As part of programs to study the variability of substellar objects, JWST is expected to observe hundreds of exoplanets, free-floating planets, and brown dwarfs, yielding a vast reservoir of light curves well-suited for the search for small transiting companions. Consequently, the method presented in this work provides a robust framework for leveraging those light curves providing a path for constraining the occurrence rate of terrestrial companions in a host regime where no such detections currently exist.

\software{{\tt dynesty}  \citep{speagle2020,koposov2022}}

\begin{acknowledgments}
This work is based on observations made with the NASA/ESA/CSA James Webb Space Telescope. The data were obtained from the Mikulski Archive for Space Telescopes at the Space Telescope Science Institute, which is operated by the Association of Universities for Research in Astronomy, Inc., under NASA contract NAS 5-03127 for JWST. These observations are associated with program \#2327.

This material is based upon work supported by the National Science Foundation Graduate Research Fellowship under Grant No. 2240310. Any opinion, findings, and conclusions or recommendations expressed in this material are those of the authors and do not necessarily reflect the views of the National Science Foundation.
\end{acknowledgments}

%

\vspace{5mm}





\appendix

\begin{figure}[!b]
\begin{centering}
\includegraphics[width=14.7cm]{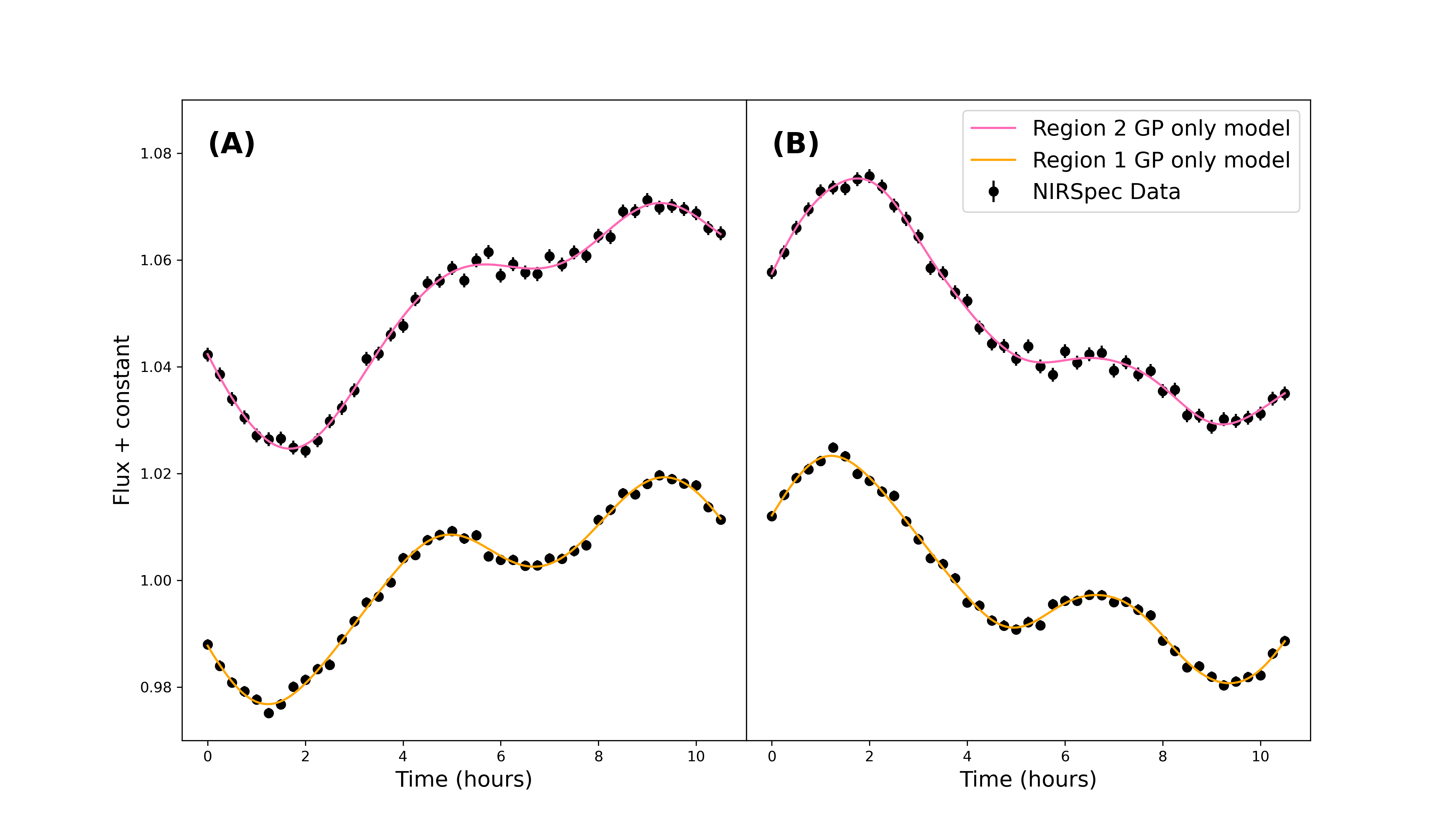}
\caption{The original light curves of WISE 0855 built from the two different wavelength regions are shown in black. The solid lines represent the fit of the GP only model. Panel A represents the original, unflipped version with no transit injections and Panel B is the flipped version of the data with the corresponding GP only model fits.}
\label{noinjgp}
\end{centering}
\end{figure}

\begin{figure}
\begin{centering}
\includegraphics[width=14.7cm]{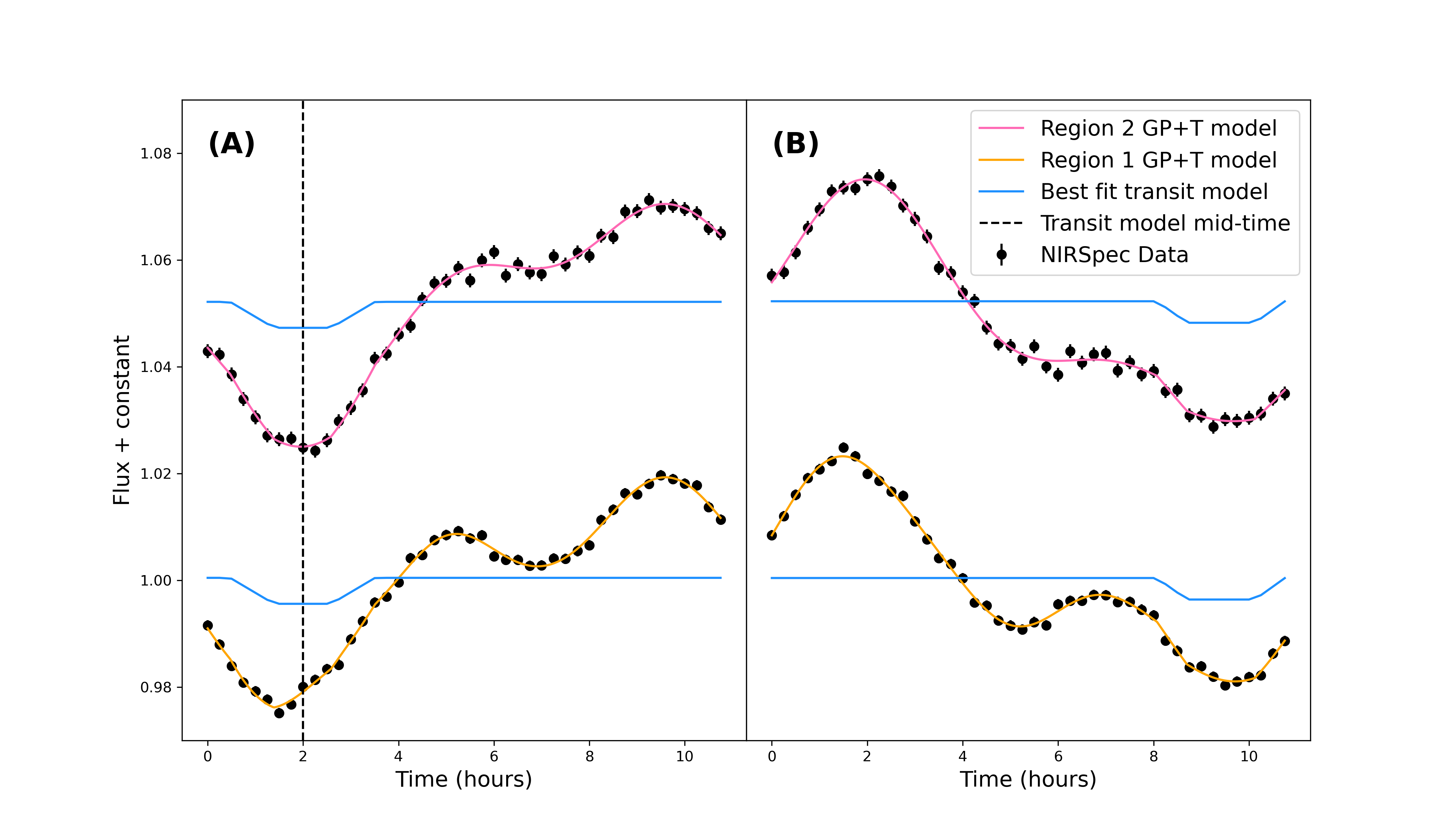}
\caption{The original light curves of WISE 0855 built from the two different wavelength regions are shown in black. The solid pink and orange lines represent the fit of the GP + Transit model. The dashed black line is the transit mid-time recovered by the transit model. Panel A represents the original, unflipped version with no transit injections and Panel B is the flipped version of the data with the corresponding GP + transit model fits. The blue line represents the best fit transit model as identified by the GP + transit model. In both the flipped and unflipped cases, we see the transit model is aligned with larger variability features that are expected to be intrinsic from the host.}
\label{noinjgpt}
\end{centering}
\end{figure}

\begin{figure}
\includegraphics[width=\columnwidth]{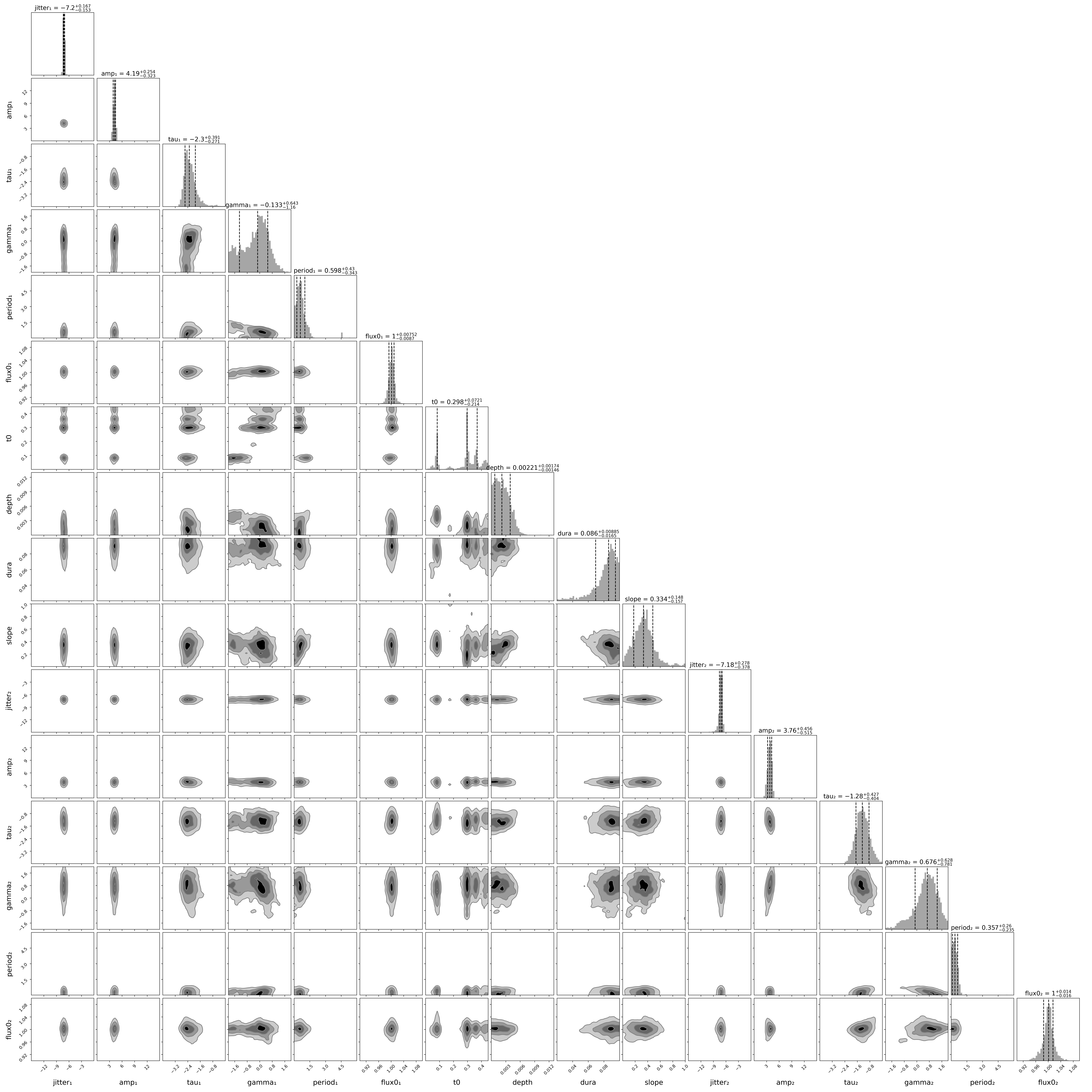}
\caption{Posterior parameter distributions of the GP + transit model results for the no-injection case on the unflipped light curves. }
\label{reggpt}
\end{figure}

\begin{figure}
\includegraphics[width=\columnwidth]{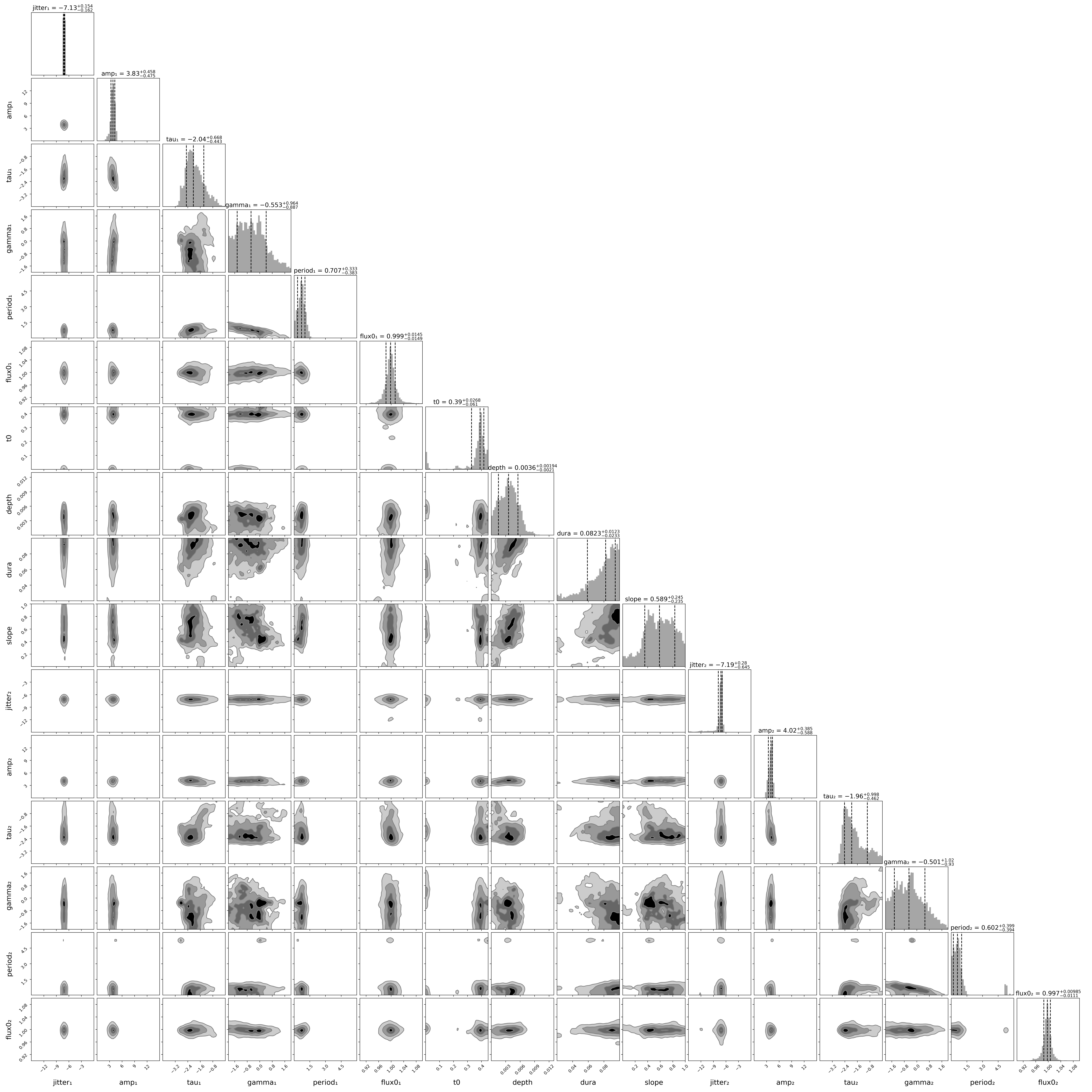}
\caption{Posterior parameter distributions of the GP + transit model results for the no-injection case on the flipped light curves. }
\label{flippedgpt}
\end{figure}

\begin{figure}
\includegraphics[width=\columnwidth]{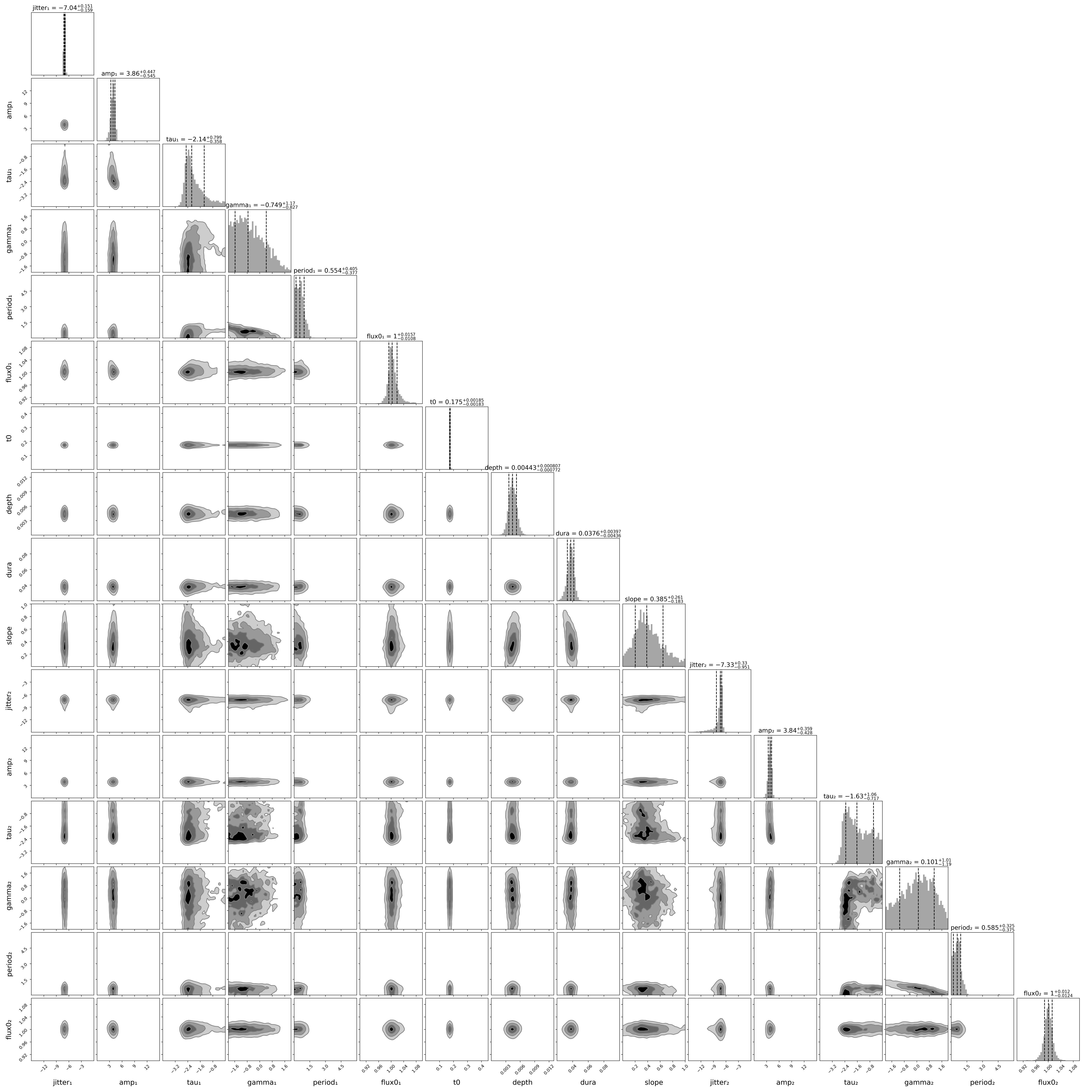}
\caption{Posterior parameter distributions of the GP + transit model results for the example injection described in section \ref{injandretsection} on the unflipped light curves. The transit specific parameters are the transit Mid-time, Depth, Duration, and Slope. In this case, all but the transit slope are more constrained than the results of both the unflipped and flipped no-injection cases. }
\label{exgpt}
\end{figure}

\begin{figure}
\includegraphics[width=\columnwidth]{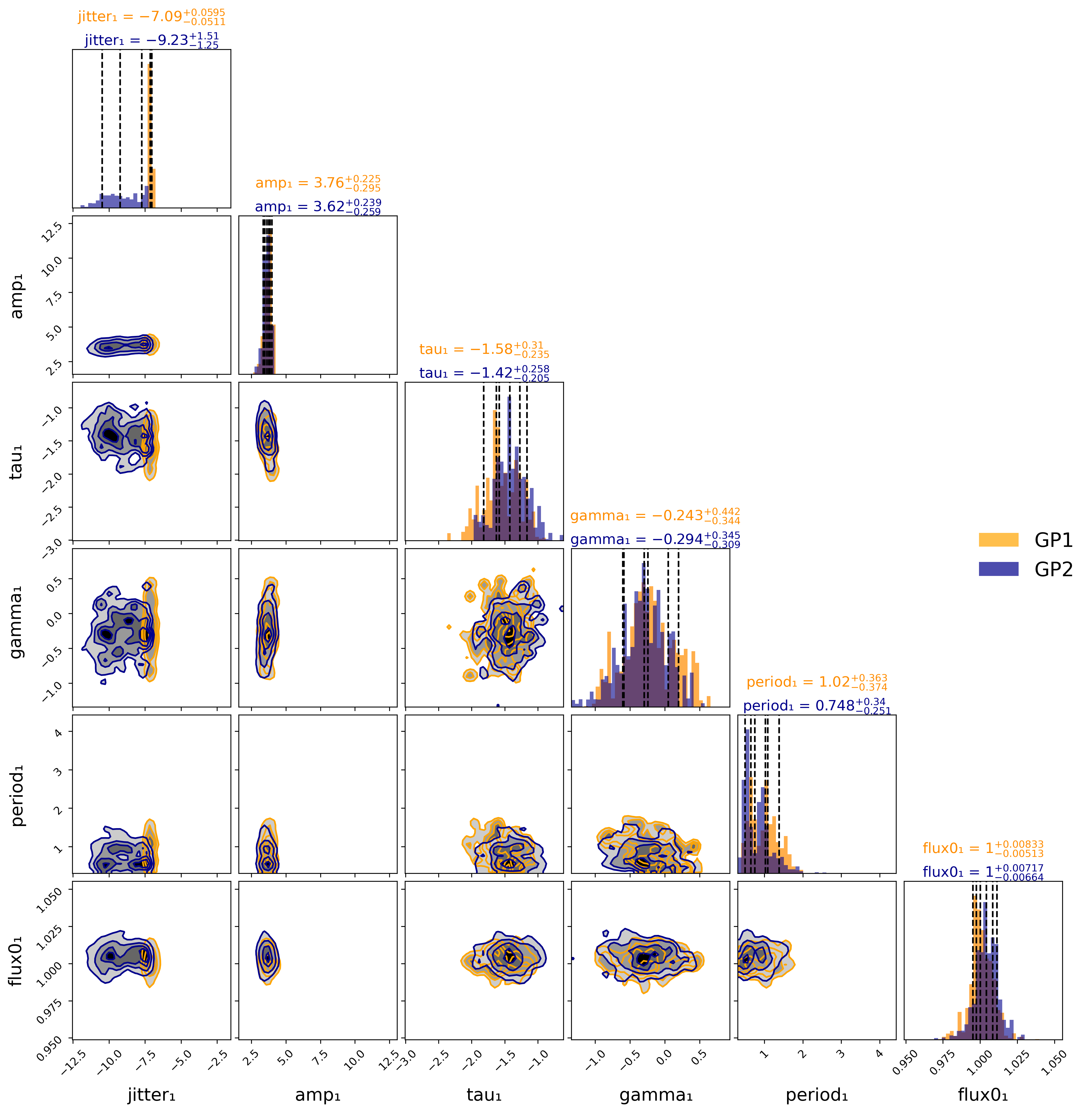}
\caption{Posterior parameter distributions of the GP only models for the the no-injection case on the unflipped light curves. Both regions are fit independently, and their results are overlayed in orange and blue for region 1 and region 2, respectively. }
\label{reggps}
\end{figure}

\begin{figure}
\includegraphics[width=\columnwidth]{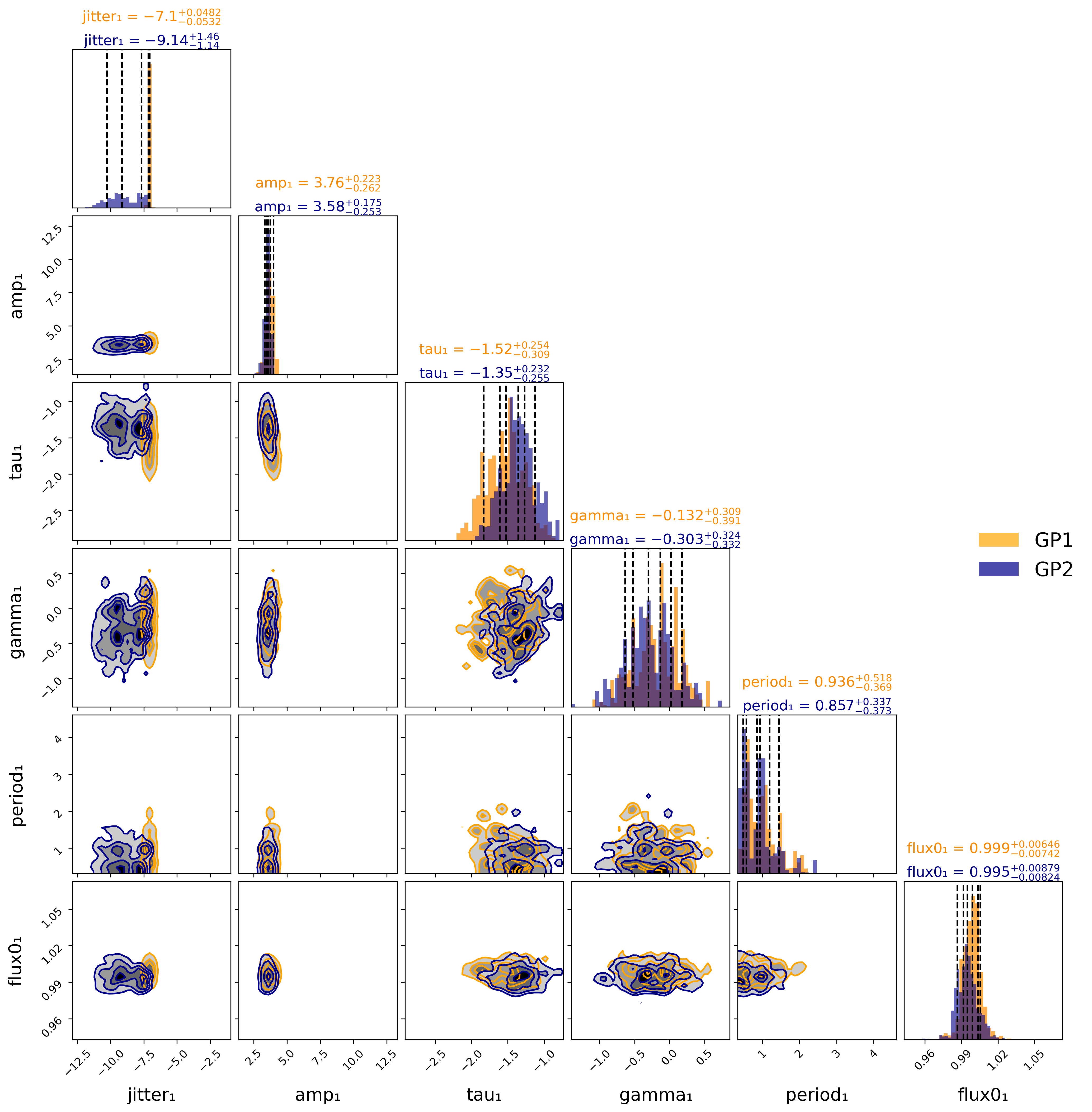}
\caption{Posterior parameter distributions of the GP only models for the the no-injection case on the flipped light curves. Both regions are fit independently, and their results are overlayed in orange and blue for region 1 and region 2, respectively. }
\label{flippedgps}
\end{figure}

\begin{figure}
\includegraphics[width=\columnwidth]{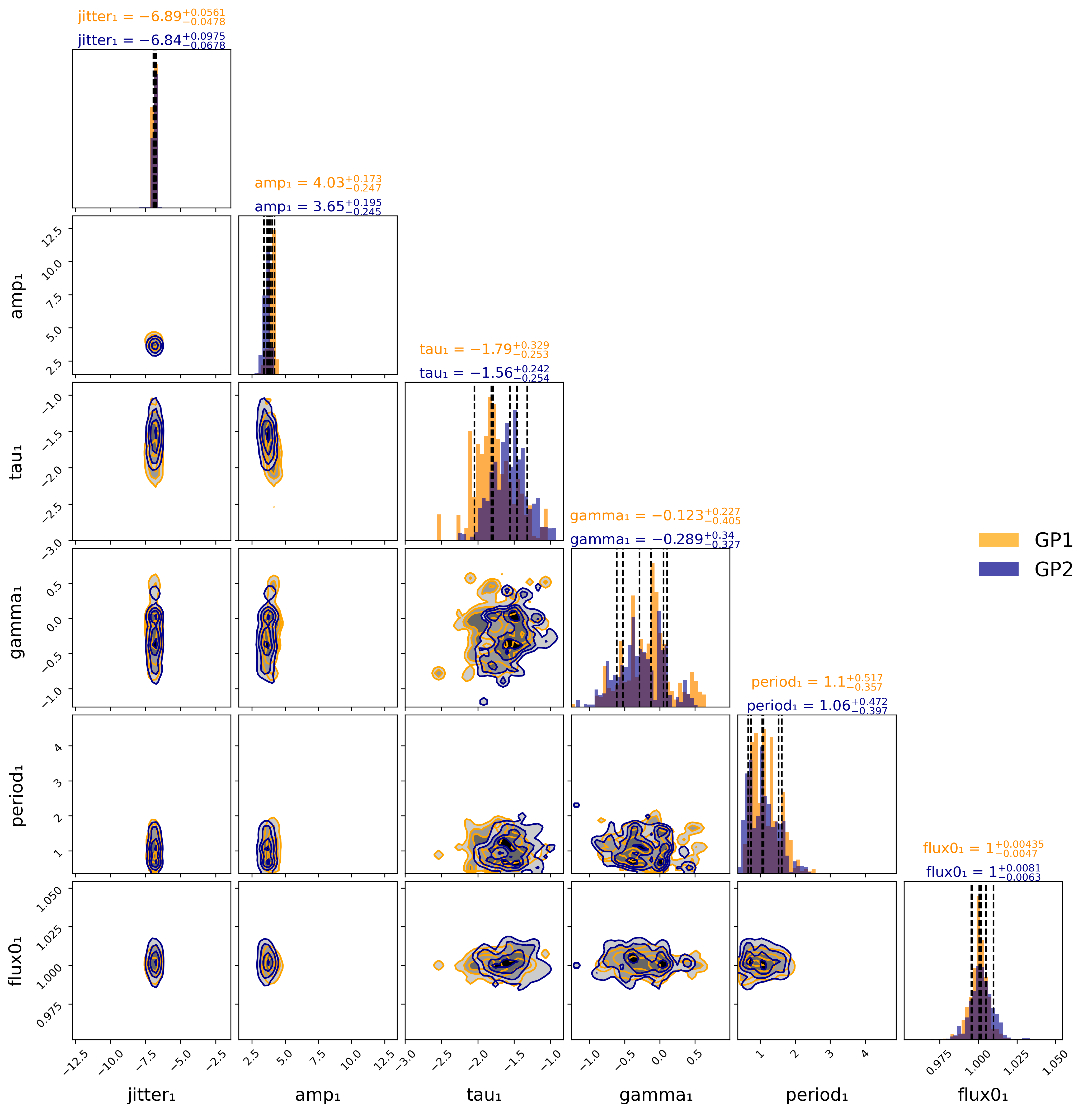}
\caption{Posterior parameter distributions of the GP only models for the the example case described in section \ref{injandretsection}. Both regions are fit independently, and their results are overlayed in orange and blue for region 1 and region 2, respectively. }
\label{exgp}
\end{figure}

Figure \ref{noinjgp} and Figure \ref{noinjgpt} show the results of the GP only and GP + transit models, respectively, for our transit search. Figure \ref{noinjgp} demonstrates the fit of the GP model to the 2 light curve regions and the corresponding flipped data. Figure \ref{noinjgpt} shows the results of the GP + transit models for the 2 light curve regions with the recovered transit shown in blue, and the transit mid-time represented with a dashed line. Figures \ref{reggpt}-\ref{exgp} show the corresponding GP + transit and GP only corner plots for each no-injection case, and the example case described in section \ref{injandretsection}.

\twocolumngrid

\pagebreak
\bibliography{bib}
\bibliographystyle{aasjournal}



\end{document}